\documentclass[twocolumn]{aastex631}
\usepackage{apjfonts}
\usepackage{graphicx}		
\usepackage{color}
\usepackage{hyperref}
\usepackage{amsmath}
\usepackage{float}		
\usepackage{amssymb}
\usepackage{textcomp}
\usepackage{tabularx}
\usepackage{gensymb}
\usepackage[procnames]{listings}
\usepackage{tabto}
\usepackage{paralist} 
\usepackage{mathtools}
\usepackage[normalem]{ulem}
\usepackage{xcolor}
\DeclareGraphicsExtensions{.pdf,.png,.jpeg}
\DeclareUnicodeCharacter{2212}{-}
\usepackage{multirow}
\usepackage{xspace}

\usepackage[encapsulated]{CJK}
\usepackage{ucs}
\newcommand{\cntext}[1]{\begin{CJK}{UTF8}{gbsn}#1\end{CJK}}

\def\apjl{ApJL}
\def\apjs{ApJS}
\def\ao{Appl.~Opt.}
\def\aap{A\&A}
\begin{document}

\newcommand{\janine}{J1924--2914\xspace}
\nocite{PaperI}
\nocite{PaperII}
\nocite{PaperIII}
\nocite{PaperIV}
\nocite{PaperV}
\nocite{PaperVI}

\newcommand{\mw}[1]{\textcolor{teal}{MW: #1}}
\newcommand{\ac}[1]{\textcolor{purple}{AC: #1}}
\newcommand{\lb}[1]{\textcolor{blue}{LB: #1}}
\newcommand{\mdj}[1]{\textcolor{olive}{MDJ: #1}}
\newcommand{\si}[1]{\textcolor{magenta}{SI: #1}}
\newcommand{\monika}[1]{\textcolor{orange}{MM: #1}}
\newcommand{\tk}[1]{\textcolor{violet}{TPK: #1}}

\newcommand{\rev}[1]{\textcolor{red}{\textbf{#1}}}

\newcommand\sbullet[1][.5]{\mathbin{\vcenter{\hbox{\scalebox{#1}{$\bullet$}}}}}

\author[0000-0002-5297-921X]{Sara Issaoun}
\affiliation{Center for Astrophysics $|$ Harvard \& Smithsonian, 60 Garden Street, Cambridge, MA 02138, USA}
\affiliation{Department of Astrophysics, Institute for Mathematics, Astrophysics and Particle
Physics (IMAPP), Radboud University, P.O. Box 9010, 6500 GL Nijmegen, The Netherlands}
\affiliation{NASA Hubble Fellowship Program, Einstein Fellow}
\email{sara.issaoun@cfa.harvard.edu}

\author[0000-0002-8635-4242]{Maciek Wielgus}
\affiliation{Max-Planck-Institut f\"ur Radioastronomie, Auf dem H\"ugel 69, D-53121 Bonn, Germany}
\email{maciek.wielgus@gmail.com}


\author[0000-0001-6158-1708]{Svetlana Jorstad}
\affiliation{Institute for Astrophysical Research, Boston University, 725 Commonwealth Ave., Boston, MA 02215, USA}

\author[0000-0002-4892-9586]{Thomas P. Krichbaum}
\affiliation{Max-Planck-Institut f\"ur Radioastronomie, Auf dem H\"ugel 69, D-53121 Bonn, Germany}

\author[0000-0002-9030-642X]{Lindy Blackburn}
\affiliation{Black Hole Initiative at Harvard University, 20 Garden Street, Cambridge, MA 02138, USA}
\affiliation{Center for Astrophysics $|$ Harvard \& Smithsonian, 60 Garden Street, Cambridge, MA 02138, USA}

\author[0000-0001-8685-6544]{Michael Janssen}
\affiliation{Max-Planck-Institut f\"ur Radioastronomie, Auf dem H\"ugel 69, D-53121 Bonn, Germany}

\author[0000-0001-6337-6126]{Chi-kwan Chan}
\affiliation{Steward Observatory and Department of Astronomy, University of Arizona, 933 N. Cherry Ave., Tucson, AZ 85721, USA}
\affiliation{Data Science Institute, University of Arizona, 1230 N. Cherry Ave., Tucson, AZ 85721, USA}

\author[0000-0002-5278-9221]{Dominic W. Pesce}
\affiliation{Center for Astrophysics $|$ Harvard \& Smithsonian, 60 Garden Street, Cambridge, MA 02138, USA}
\affiliation{Black Hole Initiative at Harvard University, 20 Garden Street, Cambridge, MA 02138, USA}

\author[0000-0003-4190-7613]{Jos\'e L. G\'omez}
\affiliation{Instituto de Astrof\'{\i}sica de Andaluc\'{\i}a-C\'{\i}SIC, Glorieta de la Astronom\'{\i}a s/n, E-18008 Granada, Spain}

\author[0000-0002-9475-4254]{Kazunori Akiyama}
\affiliation{Massachusetts Institute of Technology Haystack Observatory, 99 Millstone Road, Westford, MA 01886, USA}
\affiliation{National Astronomical Observatory of Japan, 2-21-1 Osawa, Mitaka, Tokyo 181-8588, Japan}
\affiliation{Black Hole Initiative at Harvard University, 20 Garden Street, Cambridge, MA 02138, USA}

% ...

\title{Resolving the inner parsec of the blazar J1924--2914 with the Event Horizon Telescope}

\shorttitle{Resolving the inner parsec of J1924--2914}
\shortauthors{Issaoun \& Wielgus et al.}

\begin{abstract}
The blazar J1924--2914 is a primary Event Horizon Telescope (EHT) calibrator for the Galactic Center's black hole Sagittarius A$^*$. Here we present the first total and linearly polarized intensity images of this source obtained with the unprecedented 20\,$\mu$as resolution of the EHT. J1924--2914 is a very compact flat-spectrum radio source with strong optical variability and polarization. In April 2017 the source was observed quasi-simultaneously with the EHT (April 5-11), the Global Millimeter VLBI Array (April 3), and the Very Long Baseline Array (April 28), giving a novel view of the source at four observing frequencies, 230, 86, 8.7, and 2.3\,GHz. These observations probe jet properties from the subparsec to 100-parsec scales. We combine the multi-frequency images of J1924--2914 to study the source morphology. We find that the jet exhibits a characteristic bending, with a gradual clockwise rotation of the jet projected position angle of about 90 degrees between 2.3 and 230\,GHz. Linearly polarized intensity images of J1924--2914 with the extremely fine resolution of the EHT provide evidence for ordered toroidal magnetic fields in the blazar compact core.
\end{abstract}


\section{Introduction}\label{sec:intro}
The radio source \janine (PKS 1921--293, OV--236) is a radio-loud quasar at a redshift $z=0.353$ \citep{Wills_1981,Jones2009}. The source exhibits strong optical variability and is highly polarized \citep{Wills_1981,Pica_1988,Worrall_1990}. 
While it is extremely compact at long radio wavelengths, very long baseline interferometry (VLBI) observations at centimeter wavelengths were able to resolve a persistent core-jet structure elongated in a northern direction \citep[e.g.,][]{Preston_1989,Shen_1997,Tingay_1998,Kellermann1998}. The source is a part of the 15\,GHz Monitoring Of Jets in Active galactic nuclei with VLBA Experiments (MOJAVE) source sample\footnote{\url{http://www.physics.purdue.edu/astro/MOJAVE/sourcepages/1921-293.shtml}} and shows a prominent 10\,milli-arcsecond scale jet, see \citet{Lister2018}. Imaging and Gaussian-component model fitting of observations at frequencies between 5 and 43\,GHz conducted between 1994 and 2000 indicated a sharp bend of the inner jet from north-east to north with increasing frequency \citep{Shen_1999,Shen_2002}. Motions of individual components across multiple years were observed further downstream in the jet, but not yet in the VLBI core region on sub-milliarcsecond scales.

Early 230\,GHz observations with the prototype Event Horizon Telescope (EHT) at three geographical sites (Hawai'i, California, and Arizona) provided a first model of the resolved structure in the inner jet of \janine via model fitting of amplitudes and closure phases \citep{Lu2012}. The individual components are extended in a direction consistent with the millimeter wavelength inner jet morphology. However, without (quasi-)simultaneous multi-frequency observations, these observations alone cannot link the compact millimeter structures to the large-scale centimeter jet. Furthermore, with a very limited $(u,v)$ coverage, these observations were unable to reconstruct an image of the source and track its detailed time variability.

\begin{figure*}
    \centering
    \includegraphics[width=\linewidth]{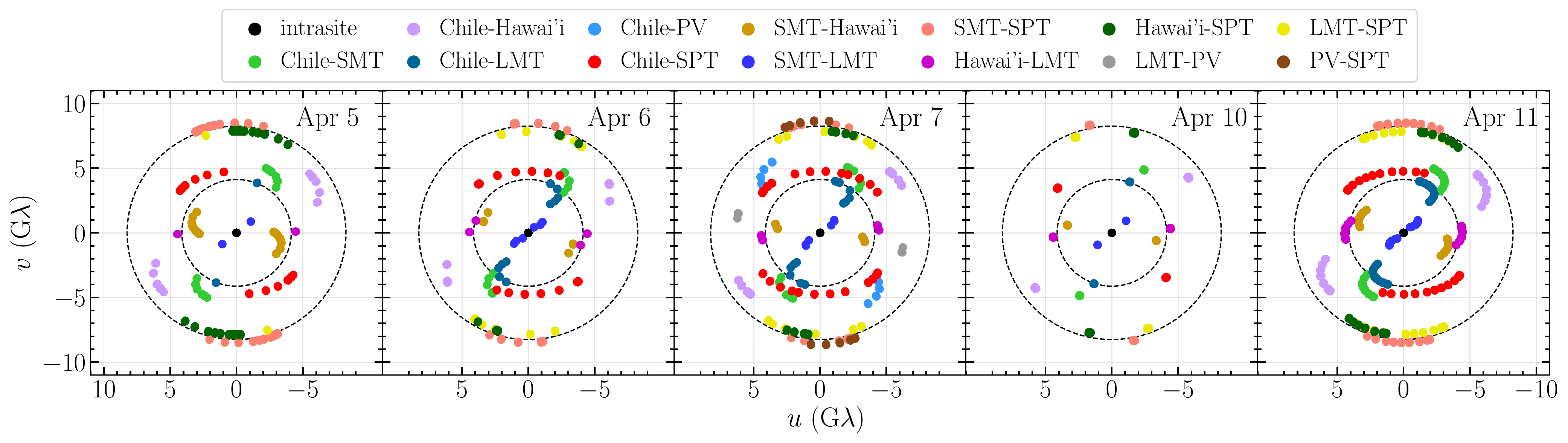}
    \caption{EHT ($u,v$) coverage on 2017 April 5, 6, 7, 10, and 11 (from left to right) for \janine. Each colored point corresponds to a single VLBI scan of $\sim5$\,min. The 2017 April 10 observations only consisted of two consecutive scans. In Chile, ALMA participated in the observations on 2017 April 6, 7, and 11, while APEX participated for all days. In Hawai'i, the SMA and JCMT both participated for all observing days. Dashed circles indicate fringe spacings characterizing the instrumental resolution of 50\,$\mu$as and 25\,$\mu$as. ``Chile'' denotes the stations ALMA and APEX. ``Hawai'i'' denotes the stations SMA and JCMT.
}
    \label{fig:eht_coverage}
\end{figure*}

\begin{figure}
    \centering
    \includegraphics[width=\linewidth]{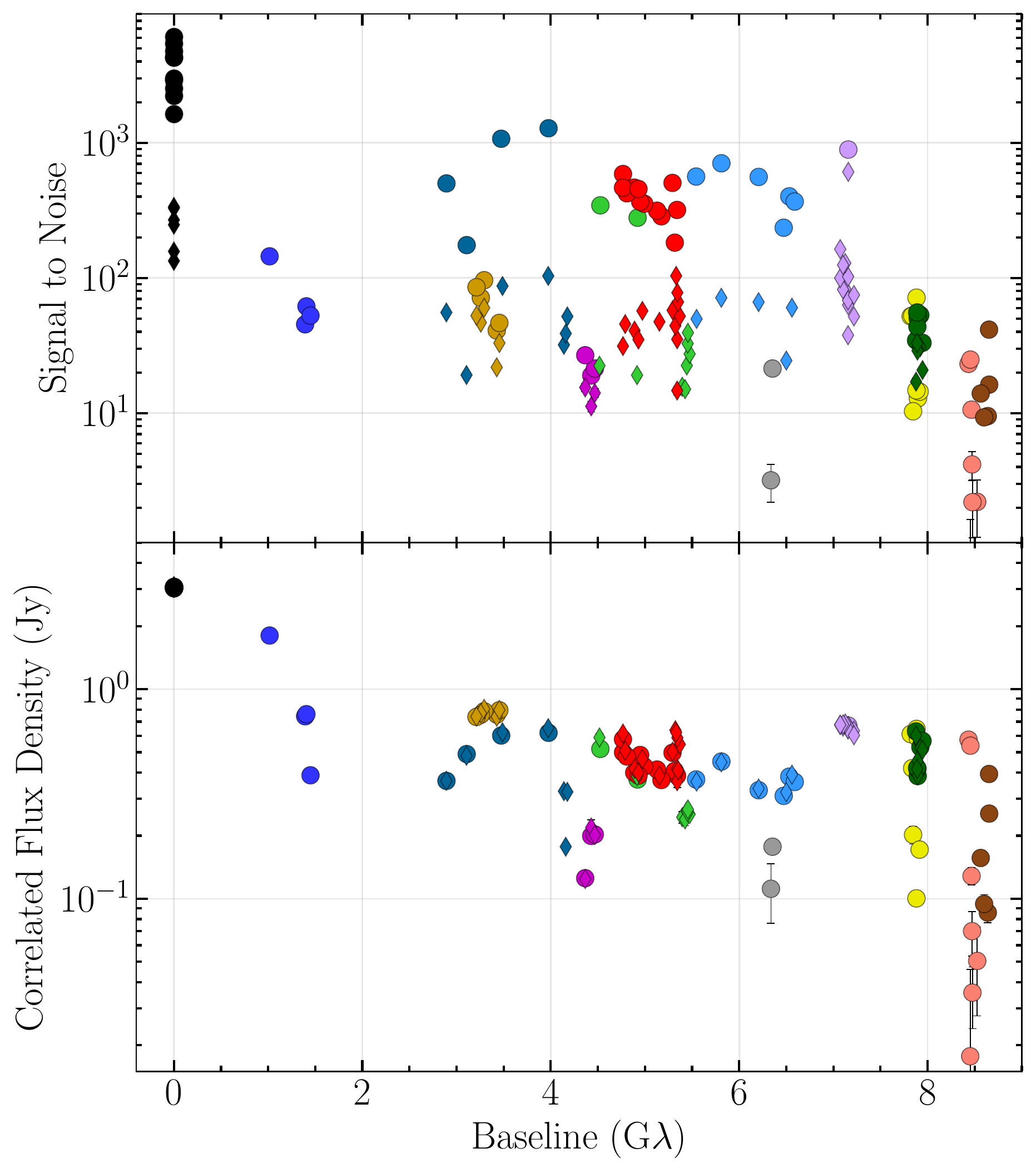}
    \caption{\textit{Top:} Signal-to-Noise ratio (S/N) for the EHT low-band observations of \janine on 2017 April 7, as a function of projected baseline length. Baselines are color-coded following Figure~\ref{fig:eht_coverage}. The circle (diamond) markers denote primary (redundant) baselines. \textit{Bottom:} Complementary plot for the visibility amplitudes after a priori calibration, in units of Jansky.}
    \label{fig:SNR_AMP}
\end{figure}

\begin{figure}
    \centering
    \includegraphics[width=\columnwidth, trim=0 0.2cm 0cm 0, clip]{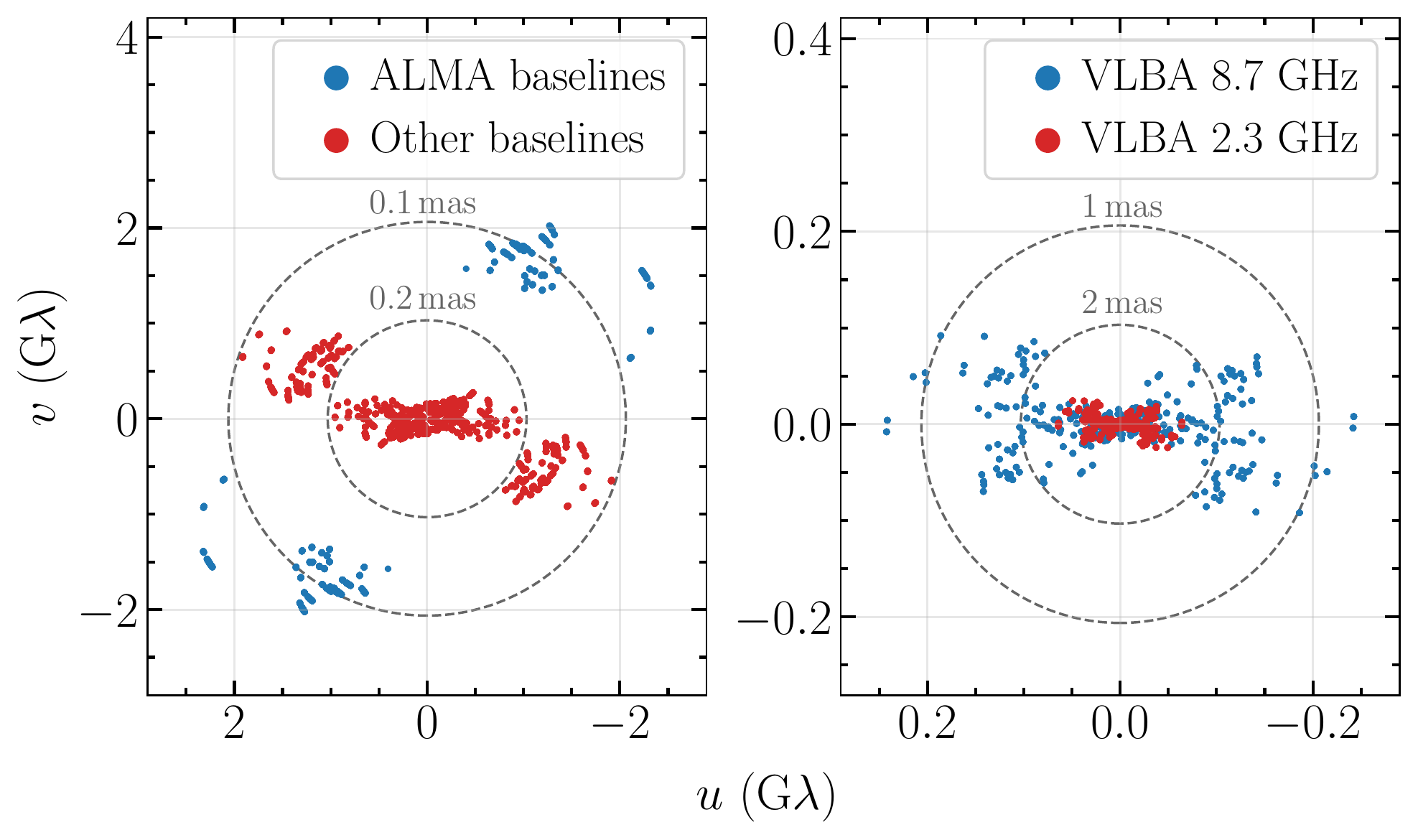}
    \caption{\textit{Left:} The $(u,v)$ coverage of the 86\,GHz observations on 2017 April 3, with the GMVA+ALMA. \textit{Right:} The $(u,v)$ coverage of the VLBA observations at 8.7\,GHz and 2.3\,GHz.
}
\label{fig:eht_coverage_3mmA}
\end{figure}

Recently, the highly sensitive Atacama Large Millimeter/submillimeter Array (ALMA) was equipped for millimeter VLBI via the ALMA Phasing Project \citep[APP;][]{Matthews_2018, Goddi2019}. In 2017, ALMA participated in its first VLBI science campaign jointly with the Global Millimeter VLBI Array (GMVA) at 86\,GHz and the EHT at 230\,GHz. In addition to its sensitivity, ALMA provides valuable north--south baselines to the predominantly east--west geometry of the GMVA. These observations enabled a first image of \janine at 86\,GHz \citep[project code MB007;][]{Issaoun_2019}. With participation of ALMA at 230\,GHz, the EHT Collaboration imaged the horizon-scale emission of M\,87$^*$ \citep{PaperI,PaperII,PaperIII,PaperIV,PaperV,PaperVI}, as well as central regions of the blazar 3C\,279 \citep{Kim2020} and the radio galaxy Cen~A \citep{Janssen2021}. The first EHT millimeter images of linearly polarized emission in M\,87$^*$ were published recently \citep{PaperVII,PaperVIII}. In this paper, we present the first total intensity and linear polarization images of \janine at 230\,GHz obtained with the EHT, and make comparisons to the near-contemporaneous GMVA results from \citet{Issaoun_2019} and Very Long Baseline Array (VLBA) observations at 2.3 and 8.7\,GHz \citep{Hunt2021}.

The EHT array achieves a resolution of $\sim$\,20\,$\mu$as. At the \janine redshift of $z = 0.353$ \citep{Jones2009}, this corresponds to a linear scale\footnote{For the cosmological parameters we have assumed $H_0 = 67.7$ km s$^{−1}$ Mpc$^{−1}$, $\Omega_m = 0.307$, and $\Omega_\Lambda = 0.693$ \citep{CosmoCalc2006,Planck2016}.} of 0.1\,pc or, in Schwarzschild radius units $R_{\rm S} = 2GM_{\sbullet[1.35]}/c^2$, about $10^3 \left(M_{\sbullet[1.35]} / 10^9 M_{\odot}  \right)^{-1} R_{\rm S}$. No robust mass estimate for \janine's central black hole was found in the literature. The EHT results reported in this paper are the highest resolution images of a blazar's linear polarization ever obtained at millimeter wavelengths, likely probing a region within the gravitational sphere of influence of the central supermassive black hole \citep[e.g.,][]{Kormedy2013}.

The paper is structured as follows. In Section~\ref{sec:obs}, we summarize the observations and data processing. We present our total intensity and polarimetric images in Section~\ref{sec:results} and discuss the theoretical implications in Section~\ref{sec:discussion}. A summary is given in Section~\ref{sec:summary}.


\section{Observations and data processing}\label{sec:obs}


\begin{figure*}
    \centering
    \includegraphics[width=0.2\textwidth]{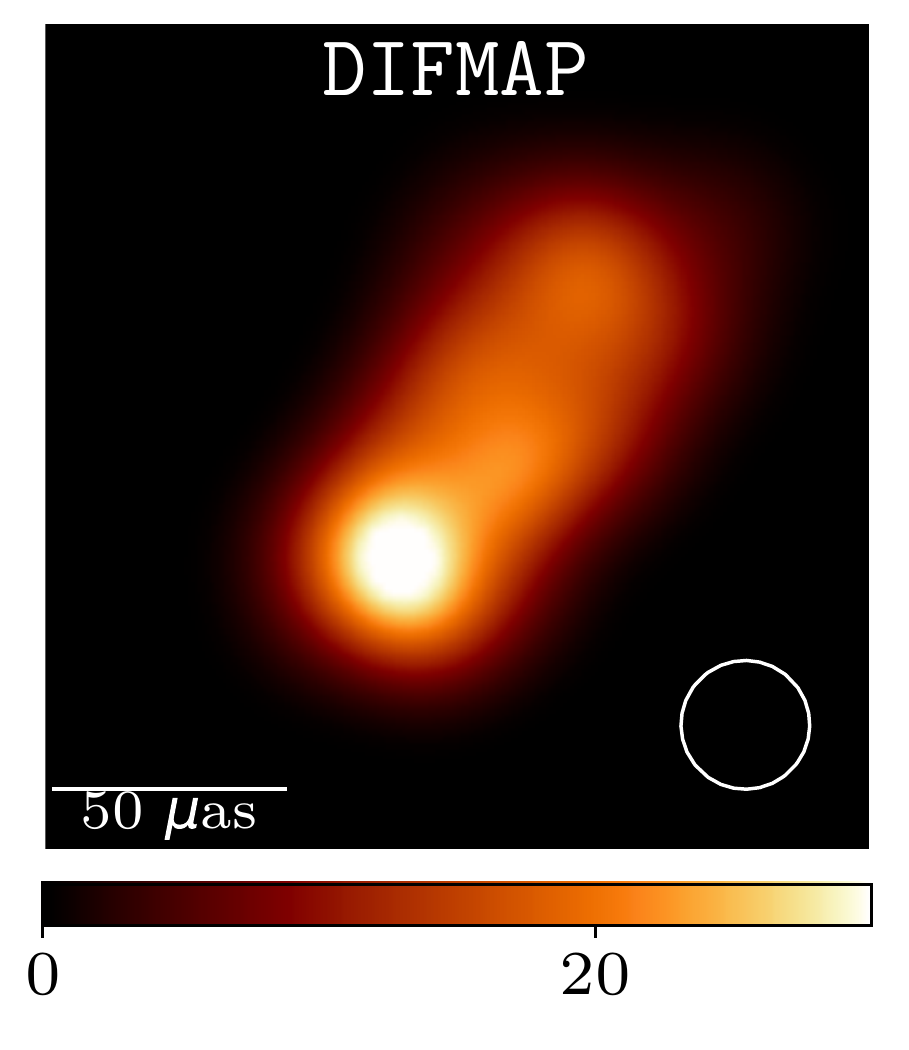}    \hspace{-0.3cm} 
    \includegraphics[width=0.2\textwidth]{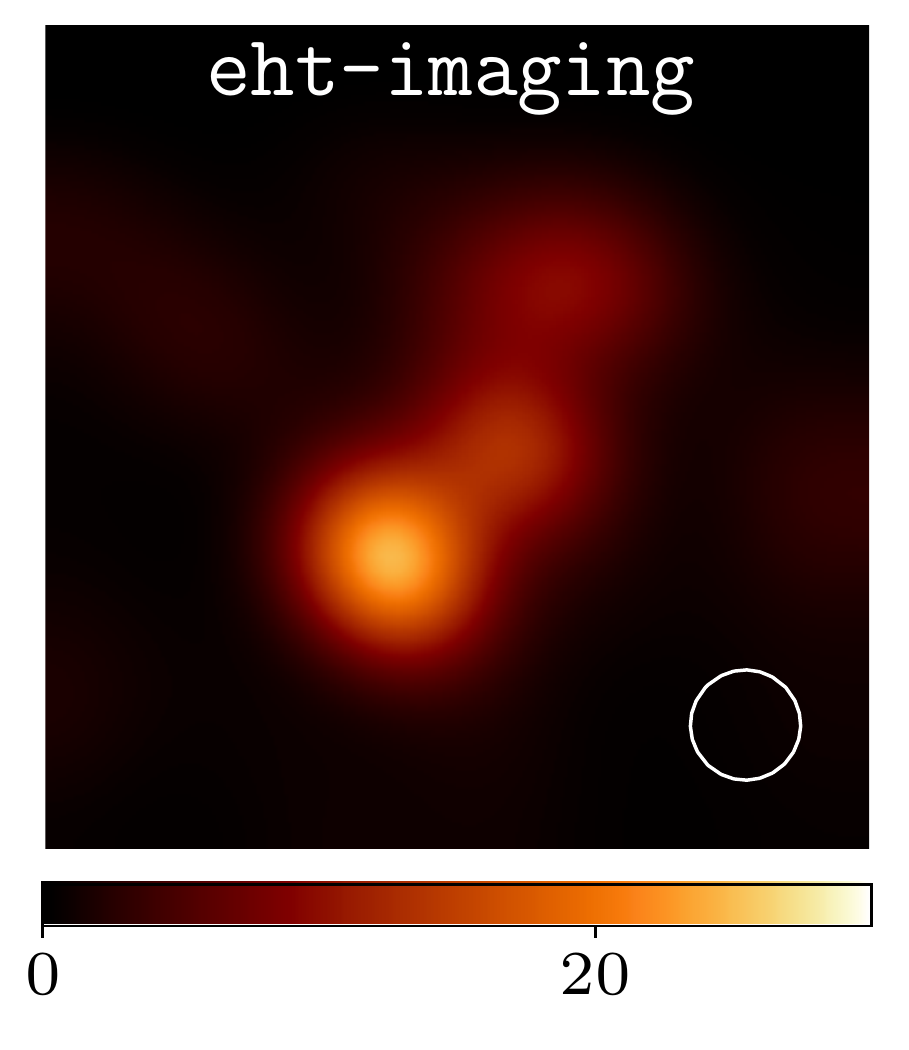} \hspace{-0.25cm}   
    \includegraphics[width=0.2\textwidth]{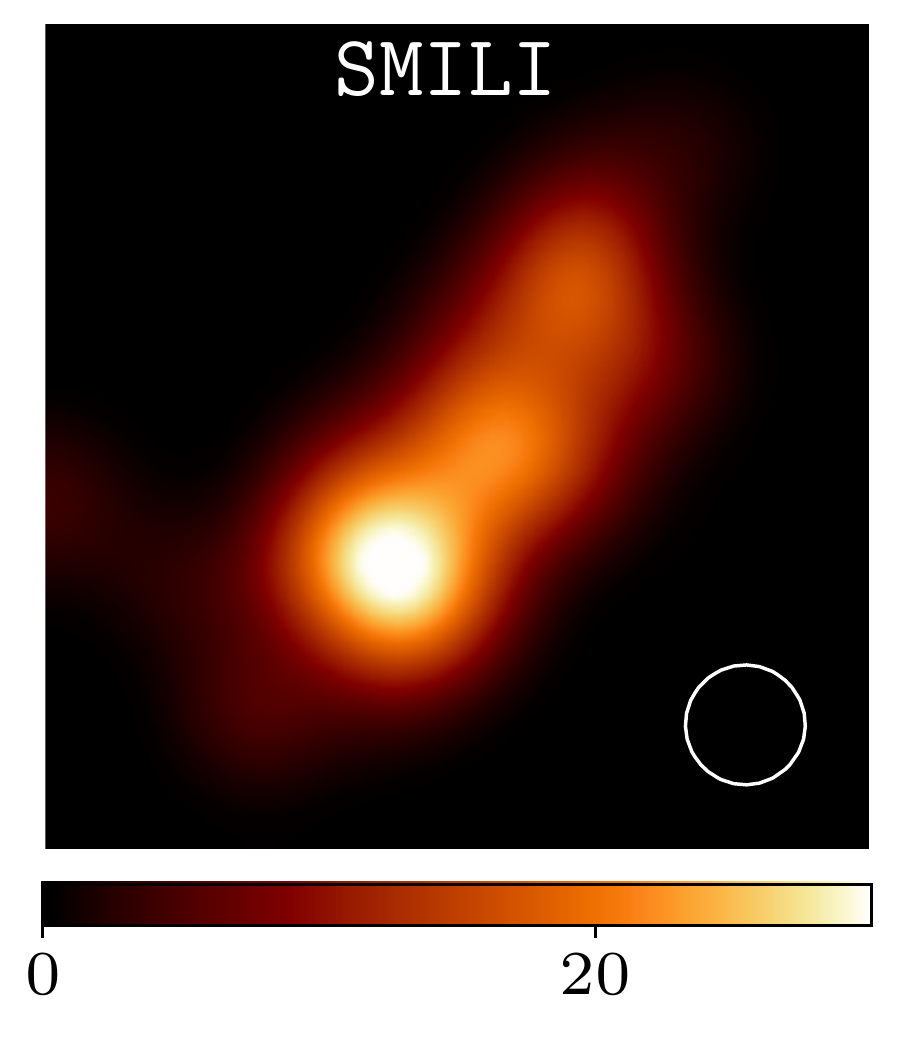}\hspace{-0.1cm}   
    \includegraphics[width=0.2\textwidth]{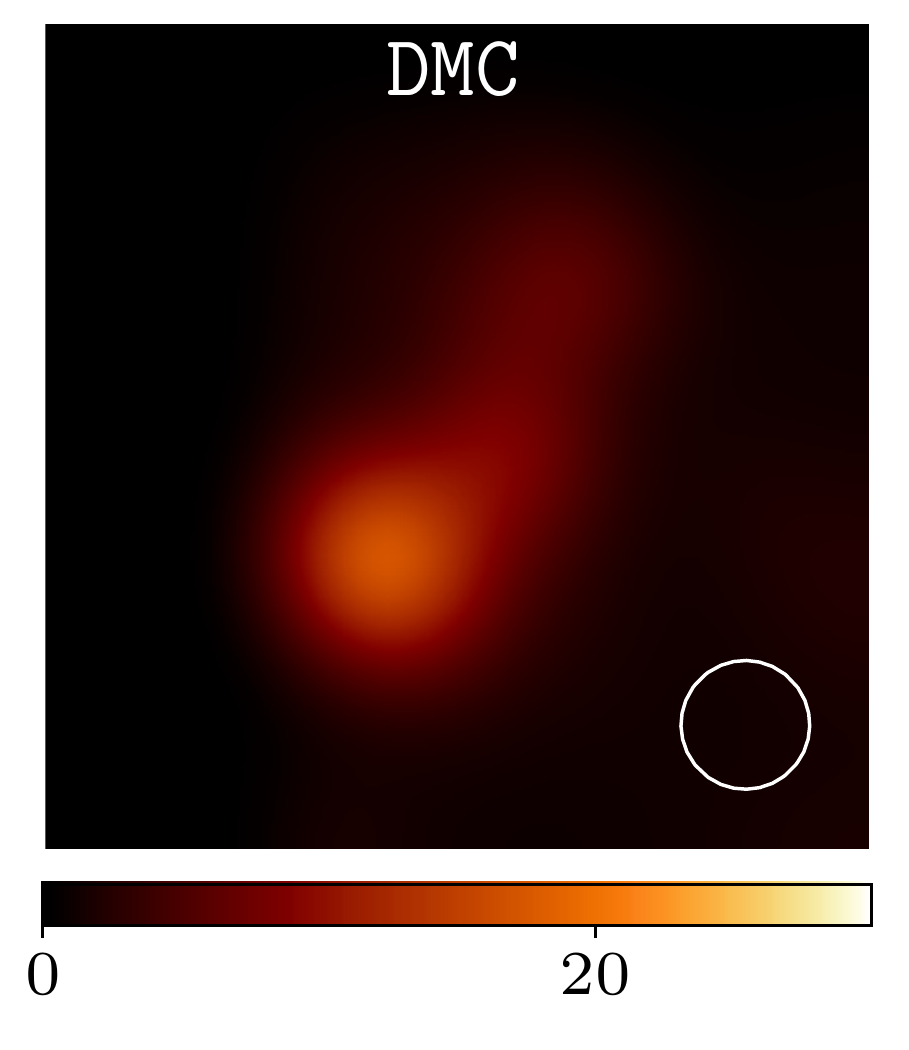} \hspace{-0.2cm}   
    \includegraphics[width=0.2\textwidth]{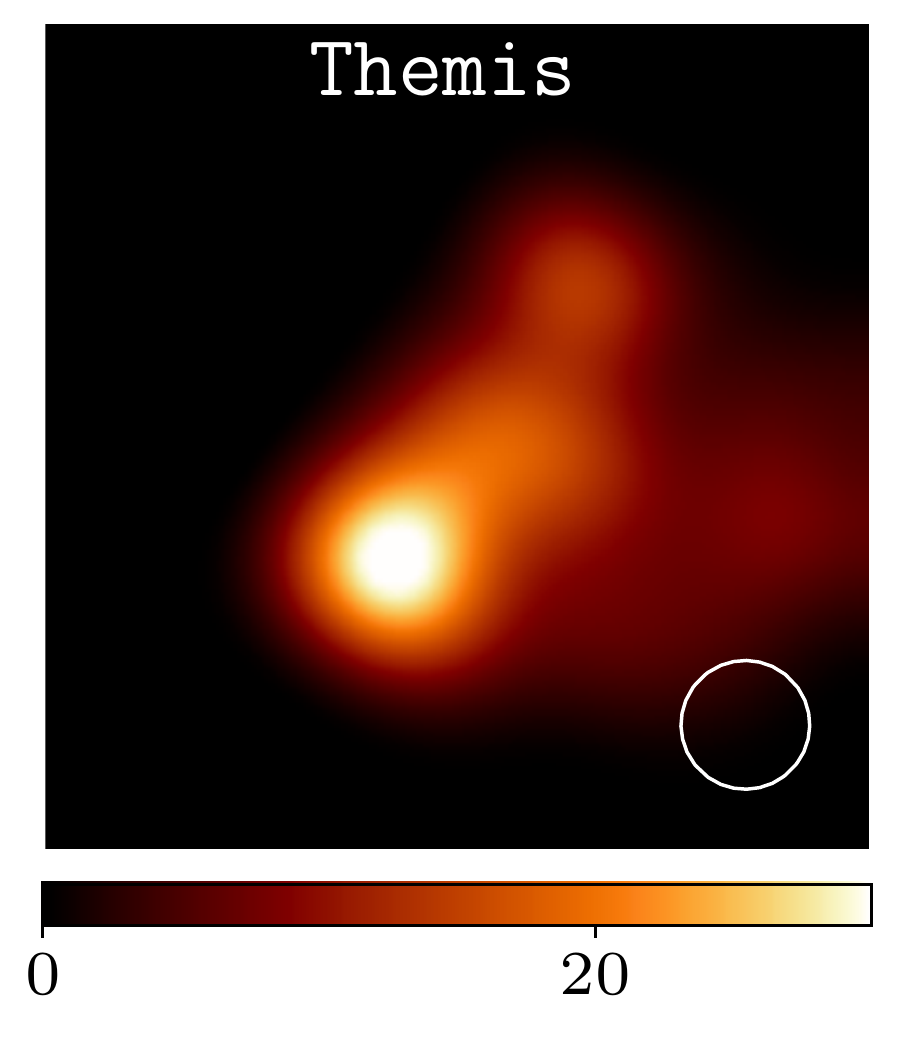}   
    \vspace{-0.5eM}\\
    {\large Brightness Temperature ($10^9$ K)}
    \caption{Representative images of \janine from the 2017 April~7 EHT observations produced using {\tt DIFMAP}, {\tt eht-imaging}, and {\tt SMILI}, and the mean posterior images from {\tt DMC} and {\tt Themis}. To simplify visual comparisons and display the images at similar resolutions, the images are restored with circular Gaussian beams of various sizes (due to varying intrinsic levels of super-resolution across softwares) for an effective resolution matching the EHT beam of 20\,$\mu$as.
}
    \label{fig:apr7_images}
\end{figure*}

\subsection{230 GHz EHT}
Observations of \janine were carried out by the EHT on 2017 April 5, 6, 7, 10, and 11, interleaved among observations of Sagittarius~A$^*$ \citep{SgraP1}, for which \janine was used as an active galactic nucleus (AGN) calibrator source, along with the blazar NRAO\,530 (Jorstad et al., in prep). Eight stations at six geographic sites took part in the observations: ALMA and the Atacama Pathfinder Experiment (APEX) telescope in Chile; the Large Millimeter Telescope Alfonso Serrano (LMT) in Mexico; the IRAM 30\,m telescope (PV) in Spain; the Submillimeter Telescope (SMT) in Arizona; the James Clerk Maxwell Telescope (JCMT) and the Submillimeter Array (SMA) in Hawai'i; and the South Pole Telescope (SPT) in Antarctica. 

The signals were recorded onto Mark6 recorders at a rate of 32\,Gbps in two $\sim$2\,GHz subbands  centered at 227.1 and 229.1\,GHz (hereafter low and high bands, respectively), using dual right-hand and left-hand circularly polarized feeds (RCP ad LCP, respectively) for all stations except ALMA and JCMT. ALMA recorded dual linear polarization, which was subsequently converted at the correlation stage to a circular basis by {\tt PolConvert} \citep{Marti_2016,Goddi2019}. The JCMT observed a single circular polarization component during the campaign (predominantly RCP for 2017 April 5 and 6 and LCP for 2017 April 7, 10, and 11). In Figure~\ref{fig:eht_coverage}, we show the resulting EHT ($u,v$) coverage for each observing day.
Good ($u,v$) coverage on 2017 April 5-7 and 11 facilitated a detailed VLBI imaging of the source at 230\,GHz.
Due to the very sparse snapshot coverage on 2017 April 10 and the static properties of the source on short timescales (see Section \ref{sec:results}), observations on 2017 April 10 were combined with those of 2017 April 11 for analysis.

After observation, the data were shipped to the MIT Haystack Observatory and the Max-Planck-Institut f\"{u}r Radioastronomie in Bonn for correlation, see \citet{PaperII} for details. The resulting data were calibrated using two separate VLBI data reduction pipelines \citep{Blackburn_2019,Janssen2019} to ensure robustness of the results. Their consistency has been studied in detail in \citet{PaperIII}, and an example of cross-pipeline comparisons is presented in Appendix~\ref{app:hops_casa}. For the science results presented in this paper we use the {\tt EHT-HOPS} pathway \citep{Blackburn_2019,PaperIII} complemented by updated postprocessing \citep[most notably with revised a priori flux density calibration;][]{SgraP2}. For further information concerning the observations, data collection, processing and validation, see \citet{PaperII,PaperIII}. 

The polarimetric calibration follows the procedures described in \citet{PaperVII}. Similarly to the M\,87$^*$ polarimetric analysis, JCMT has been flagged from the data for polarimetric imaging of \janine due to its single polarization configuration. This has no effect on the ($u,v$) coverage as all baselines to Hawai'i are fulfilled by the co-located SMA, which observed in full-polarization at all times. For the polarization leakage (D-term) calibration of the stations with a co-located site (ALMA \& APEX, SMA \& JCMT), we used the D-terms reported in Appendix D of \citet{PaperVII}, which were obtained through a robust multi-source fit to the EHT data using \texttt{polsolve} \citep{Marti2021}.
For the remaining stations (except SPT), the values adopted are derived based on the results reported in Appendix E of \citet{PaperVII} and are presented in Table~\ref{tab:dterms}. The SPT D-terms were fitted as part of analysis presented in this paper. A consistency test of the assumed leakage coefficients and constraints on the SPT D-terms are given in Appendix~\ref{app:leakage}.

\begin{table}[htp]
\caption{Leakage calibration D-terms assumed for stations without a co-located site}
\begin{center}
\tabcolsep=0.25cm
\begin{tabularx}{0.7\linewidth}{ccc}
\hline
\hline
Station & $D_{\rm R} (\%)$ & $D_{\rm L} (\%)$ \\ 
 \hline
 LMT & $2.5 + 3.5i$ & $-1.0+ 1.5i$ \\  
 SMT & $2.8 +9.0i$ & $-3.5 +10.0i$  \\ 
 PV & $-13.0 + 3.5i$ & $15.0+0.0i$   \\ 
 \hline 
\end{tabularx}
\end{center}
Note -- Constraints on SPT D-terms are discussed in Appendix~\ref{app:leakage}.
\label{tab:dterms}
\end{table}

Detections were obtained for \janine on all participating baselines of the EHT for all observing days. In Figure~\ref{fig:SNR_AMP}, we show an example of the signal-to-noise ratio (S/N) and correlated flux density on our EHT baselines for 2017 April 7 (low band), which corresponds to the observing day with the best $(u,v)$ coverage. The baselines to ALMA provide extremely high S/N of several hundreds for data averaged in 5\,min intervals, the corresponding APEX baselines offer a S/N about an order of magnitude lower.
The data shown in the bottom panel of Figure~\ref{fig:SNR_AMP} have been calibrated 
using a priori calibration information (system temperatures, antenna gains and opacities) provided for the telescopes.
The consistency of measurements on the primary (to the sensitive ALMA or SMA) and redundant (to APEX or JCMT, co-located with ALMA and SMA, respectively) baselines verifies the 10\% accuracy of the flux density calibration and indicates a good self-consistency of the data set. A network calibration procedure was additionally applied to obtain the final data set \citep{Blackburn_2019}, further improving gain calibration of sites with a co-located station by assuming a total compact flux density provided by ALMA \citep{Goddi2021}.

\begin{table*}
    \centering
    \caption{Reduced $\chi^2$ values on closure quantities and visibility amplitudes after self-calibrating to the Stokes $\mathcal{I}$ images.}
    \begin{tabular}{ccccccc}
    \hline 
    \hline 
       Date & Data Product & {\tt DIFMAP} & {\tt eht-imaging} & {\tt SMILI} & {\tt DMC} & \tt{Themis} \\
       \hline 
       \multirow{3}{*}{2017 April 5}  & $\chi^2_{\rm AMP}$ &  0.346 & 0.343 & 0.551 & 0.342 & 0.485 \\
         & $\chi^2_{\rm CP}$ &  1.211 & 1.087  & 1.338 & 1.802 & 1.264 \\
         & $\chi^2_{\rm  logCA}$ &  1.009 & 0.838 & 1.702 & 2.539 & 1.377 \\
         \hline
         \multirow{3}{*}{2017 April 6}  & $\chi^2_{\rm AMP}$ & 1.226  & 0.579 & 2.13 & 0.791  & 1.143 \\
         & $\chi^2_{\rm  CP}$ &  1.850 & 0.910 & 0.935 &1.074 & 1.350 \\
         & $\chi^2_{\rm  logCA}$ & 2.128  & 0.978 & 3.901 &1.337 & 2.026 \\
         \hline
        \multirow{3}{*}{2017 April 7}  & $\chi^2_{\rm AMP}$ & 1.018  & 0.663 &1.00 &0.545 & 1.115 \\
         & $\chi^2_{\rm  CP}$ &  2.013 & 0.674 & 2.912 &0.828 & 2.024 \\
         & $\chi^2_{\rm logCA}$ & 1.983  & 1.191 & 1.791 & 1.286 & 2.031 \\
         \hline
        \multirow{3}{*}{2017 April 10+11}  & $\chi^2_{\rm AMP}$ &  1.994 & 1.171 &1.111 &0.914 & 1.671 \\
         & $\chi^2_{\rm CP}$ & 2.895  & 1.123 & 1.437 &1.211 & 3.592 \\
         & $\chi^2_{\rm logCA}$ & 3.806  & 1.700 & 1.907 &1.708 & 3.134  \\
         \hline
         \hline 
    \end{tabular}
    
    \label{tab:chi-squares}
    NOTE -- 
    Reduced $\chi^2$ are calculated using a total error budget containing thermal noise plus an additional complex systematic error corresponding to 2\% of the observed visibility amplitude
\end{table*}

\begin{figure*}
    \centering
    \includegraphics[height=0.25\textwidth]{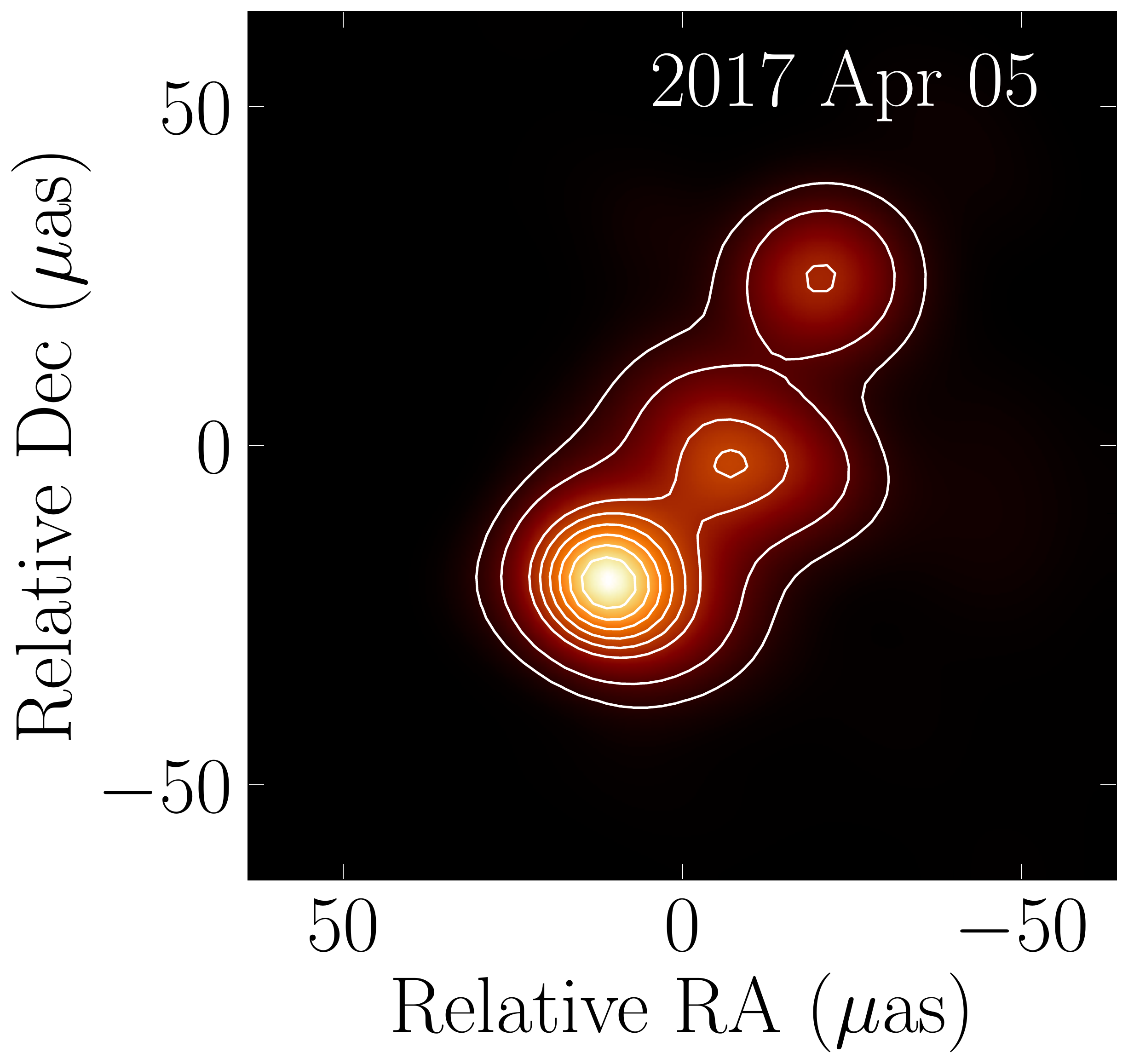} \hspace{-0.20cm}    
    \includegraphics[height=0.25\textwidth]{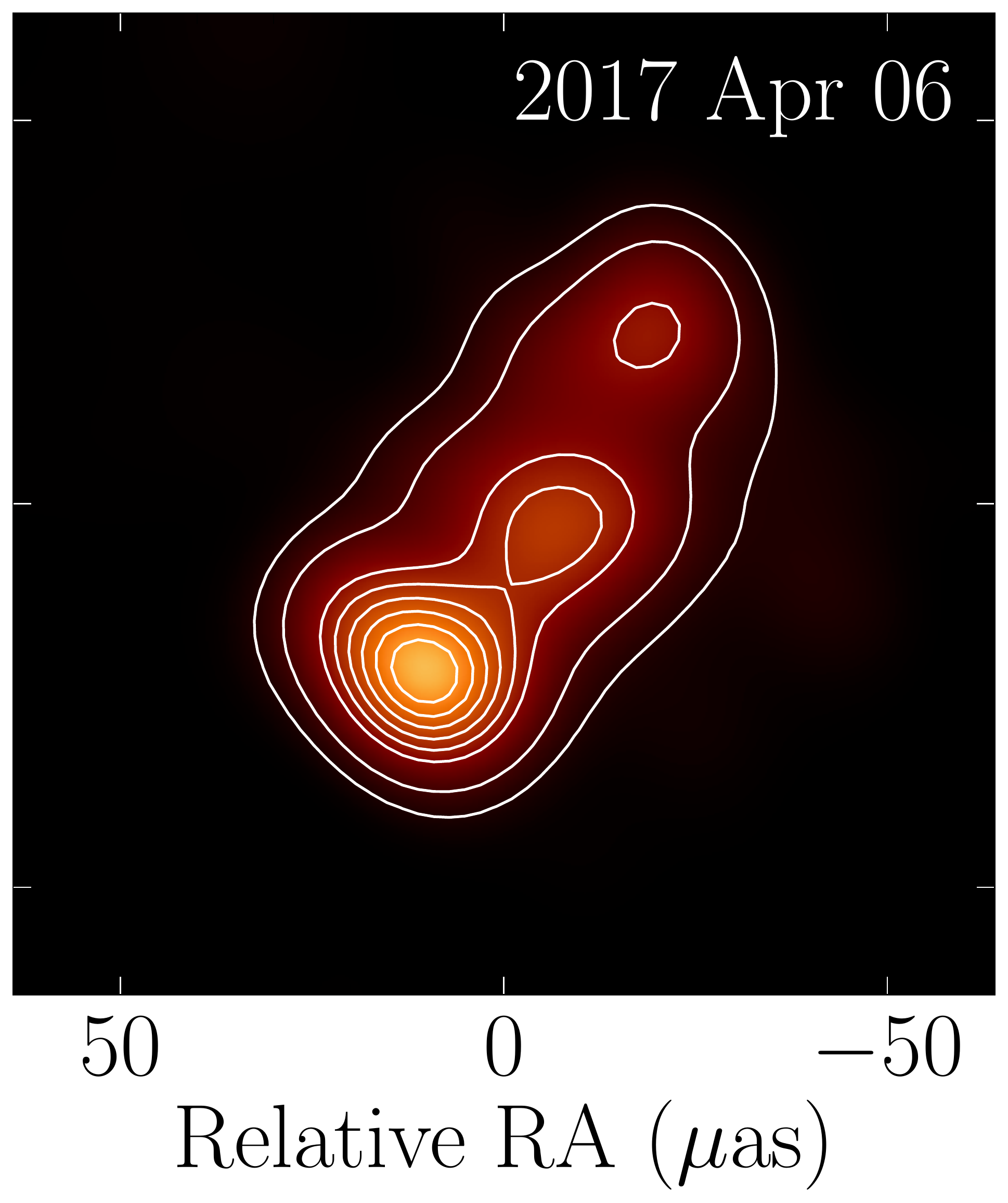}\hspace{-0.16cm}   
    \includegraphics[height=0.25\textwidth]{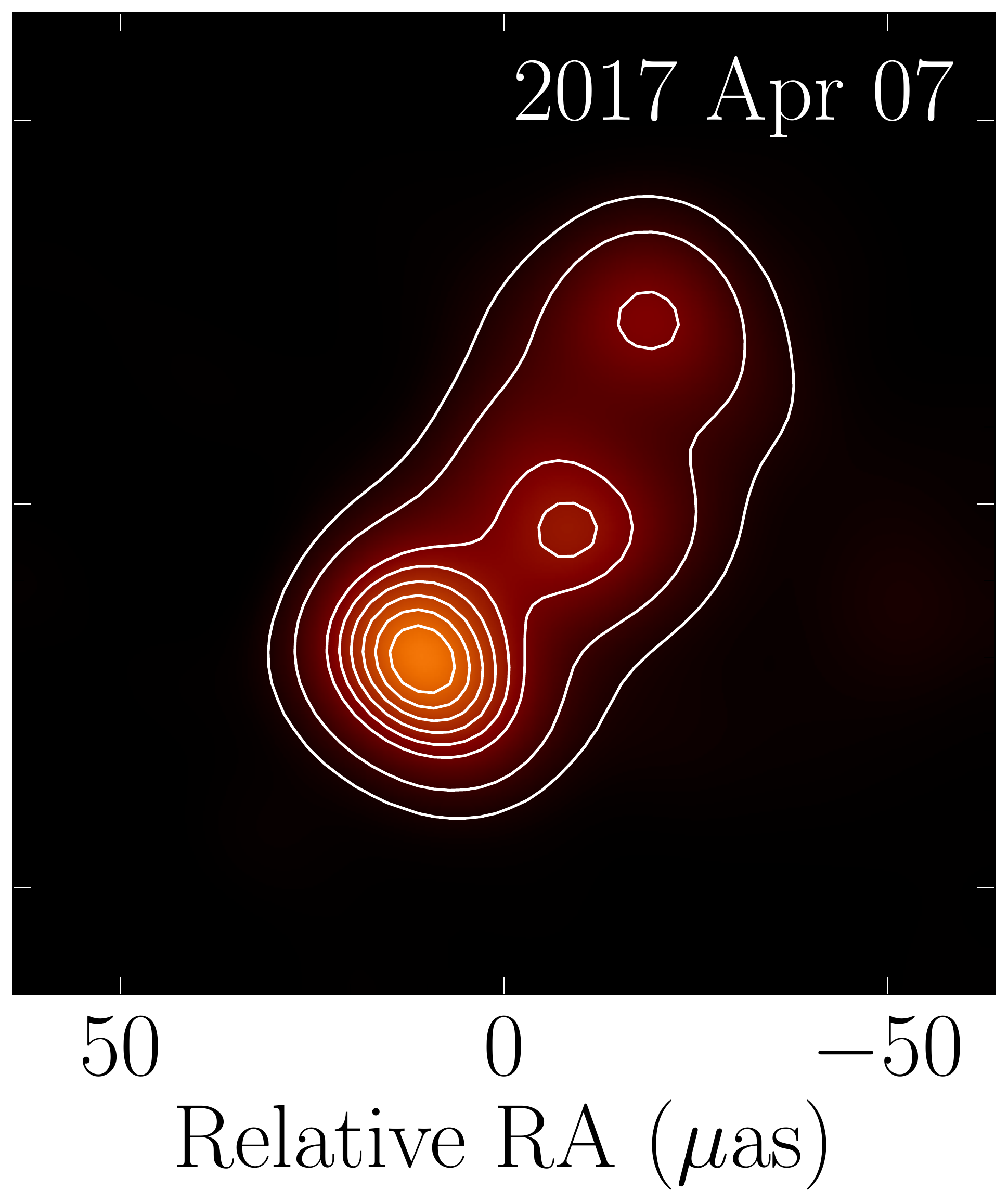}\hspace{-0.16cm}   
    \includegraphics[height=0.269\textwidth]{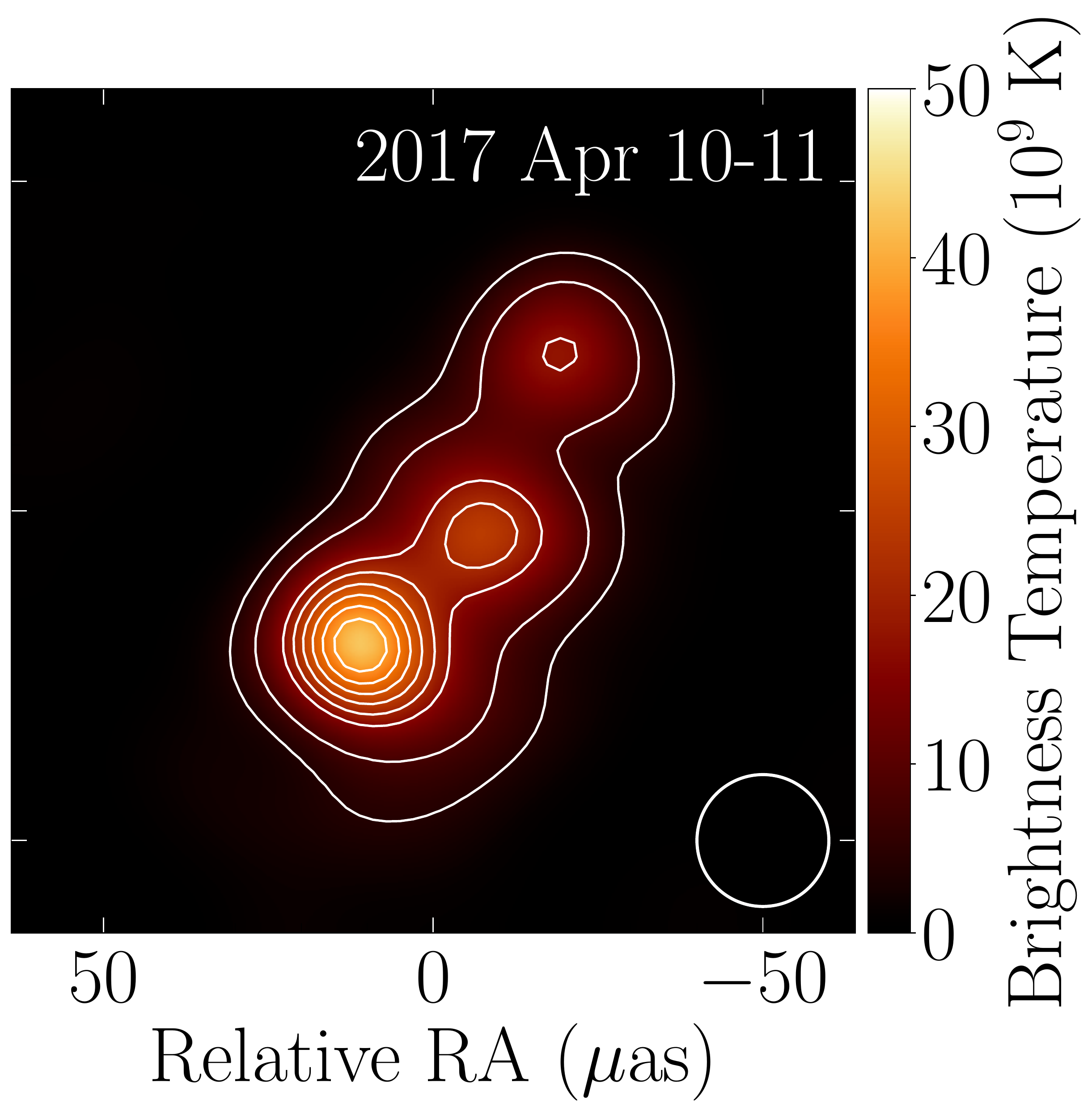}
    \caption{Method-average images of J1924--2914 from the 2017 EHT observations. The results of three imaging methods (\texttt{eht-imaging, SMILI, DIFMAP}) and two posterior exploration methods (\texttt{DMC} and \texttt{Themis}) were averaged for each of the four observing epochs. The lowest contour corresponds to 10\% of the peak intensity, with increasing contours in steps of 10\%. The CLEAN nominal beam of $20\,\mu$as FWHM is shown as a representative resolution of the average images, hereafter referred to as the EHT beam.  
}
    \label{fig:avg_images}
\end{figure*}

\subsection{86 GHz GMVA+ALMA}
Observations of \janine at 86\,GHz ($\lambda$3.5\,mm) were obtained with the GMVA+ALMA on 2017 April 3 in conjunction with observations of Sagittarius~A$^*$ published in \citet{Issaoun_2019}. The array was composed of eight VLBA antennas, the Green Bank Telescope, the Yebes 40-m telescope, the IRAM 30-m telescope, the Effelsberg 100-m telescope and ALMA. The data were recorded with a bandwidth of 256\,MHz at a data rate of 4\,Gbps over a 12\,h track, of which 8\,h included ALMA. The total recorded time on \janine was about 2 hours. In the left panel of Figure \ref{fig:eht_coverage_3mmA} we show the $(u,v)$ coverage of these observations. These observations yield images with a beam size of ($122 \times 88$)\,$\mu$as at 36\,deg. The data reduction, processing, and imaging followed a similar pathway to the EHT data, and are described in more detail in \citet{Issaoun_2019}. Analysis of this data set was challenging because of the large uncertainties in the amplitude gain calibration and low phase stability of the complex visibilities. The 86\,GHz image of \janine presented in this work was reconstructed with the {\tt eht-imaging} library using only closure quantities \citep{Chael_2016,Chael_2018} to overcome the calibration problems and was originally published in \citet{Issaoun_2019}. Linear polarization imaging at 86\,GHz did not yield robust results that could be interpreted confidently.

\subsection{2.3 and 8.7\,GHz VLBA}

Observations of \janine at 2.3 and 8.7\,GHz were carried out as part of the International Celestial Reference Frame survey with the VLBA \citep{Hunt2021}. The observations were executed in astrometric and geodetic modes, thus providing high positional accuracy. The target was observed simultaneously at the two frequencies, as is customary for geodetic observations. To reduce any variability-induced offset, we select the observation that is closest in time to the EHT observations reported here, on 2017 April 28. The target was observed by all ten stations of the VLBA, where sixteen intermediate frequency sub-bands were recorded, with four centered at 2.3\,GHz and twelve at 8.7\,GHz, for a total bandwidth of 128 and 384\,MHz at the respective frequencies. In the right panel of Figure \ref{fig:eht_coverage_3mmA} we show the $(u,v)$ coverage of these observations. Observations were recorded in right-hand circular polarization mode only, at a data rate of 2\,Gbps. For further information on observation and data calibration, refer to \citet{Hunt2021}. The calibrated data were imaged using the CLEAN algorithm implemented in the \texttt{DIFMAP} software package \citep{Shepherd_1997,Shepherd_2011}. The beam sizes are ($9.58 \times 3.58$)\,mas at $-5$\,deg and ($2.45 \times 0.94$)\,mas at $-3$\,deg for the 2.3 and 8.7\,GHz observations, respectively. We iterated the reconstruction process using a hybrid imaging loop, consisting of CLEAN and a phase self-calibration cycle, with a stopping criterion of obtaining three times the noise floor of the residual image. The final images are presented in Section~\ref{sec:multifreq} as part of the multi-frequency analysis.

\begin{figure*}
    \centering
    \includegraphics[width=\textwidth]{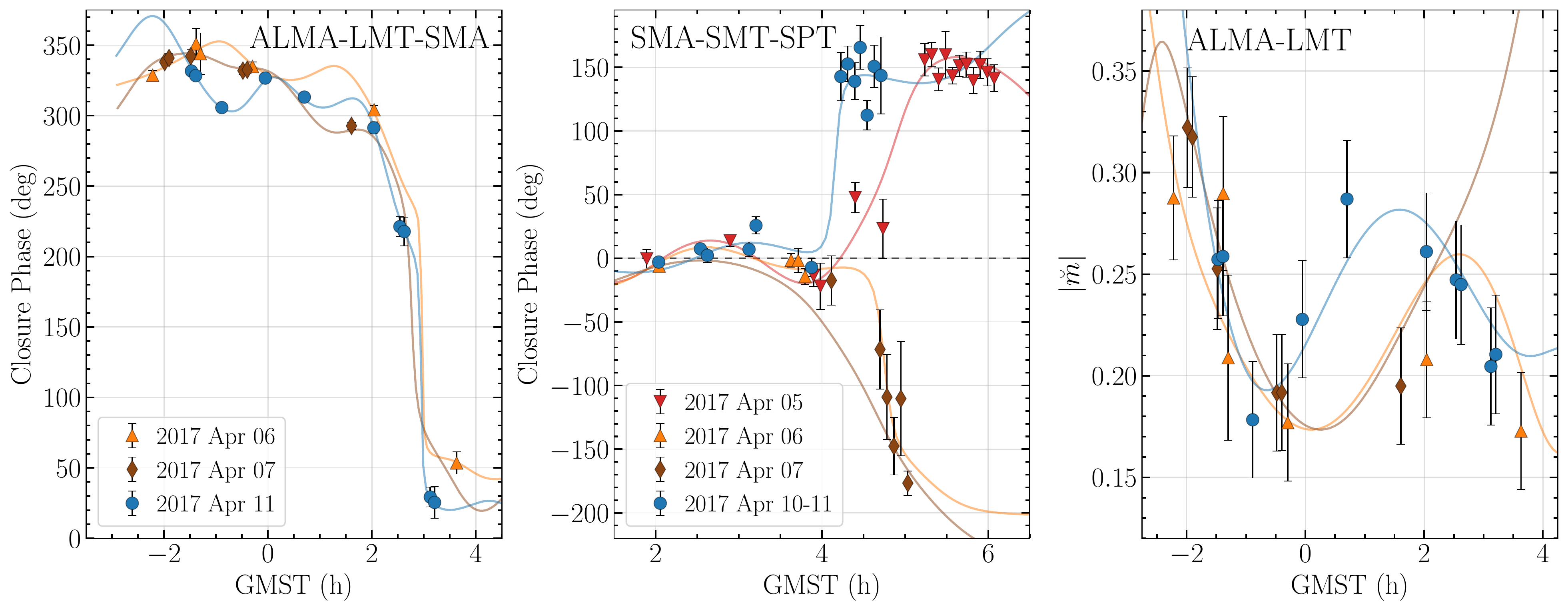}
    \caption{Consistency between observations and the RML images obtained with \texttt{eht-imaging} (continuous lines). \textit{Left}: closure phases on the sensitive ALMA--LMT--SMA triangle, exhibiting large variation related to the evolution of the triangle geometry during the Earth's rotation. \textit{Middle:} Closure phases on the very extended SMA--SMT--SPT triangle. A degeneracy between models fitted to different days can be observed, with hints of the structural evolution between days. \textit{Right:} fractional Fourier polarization (Equation \ref{eq:mbreve}) on the sensitive ALMA--LMT baseline. The errors are dominated by the systematic uncertainty corresponding to 2\,\% of the Stokes $\mathcal{I}$ visibility.}
    \label{fig:consistency}
\end{figure*}

\begin{figure}
    \centering
    \includegraphics[width=\linewidth]{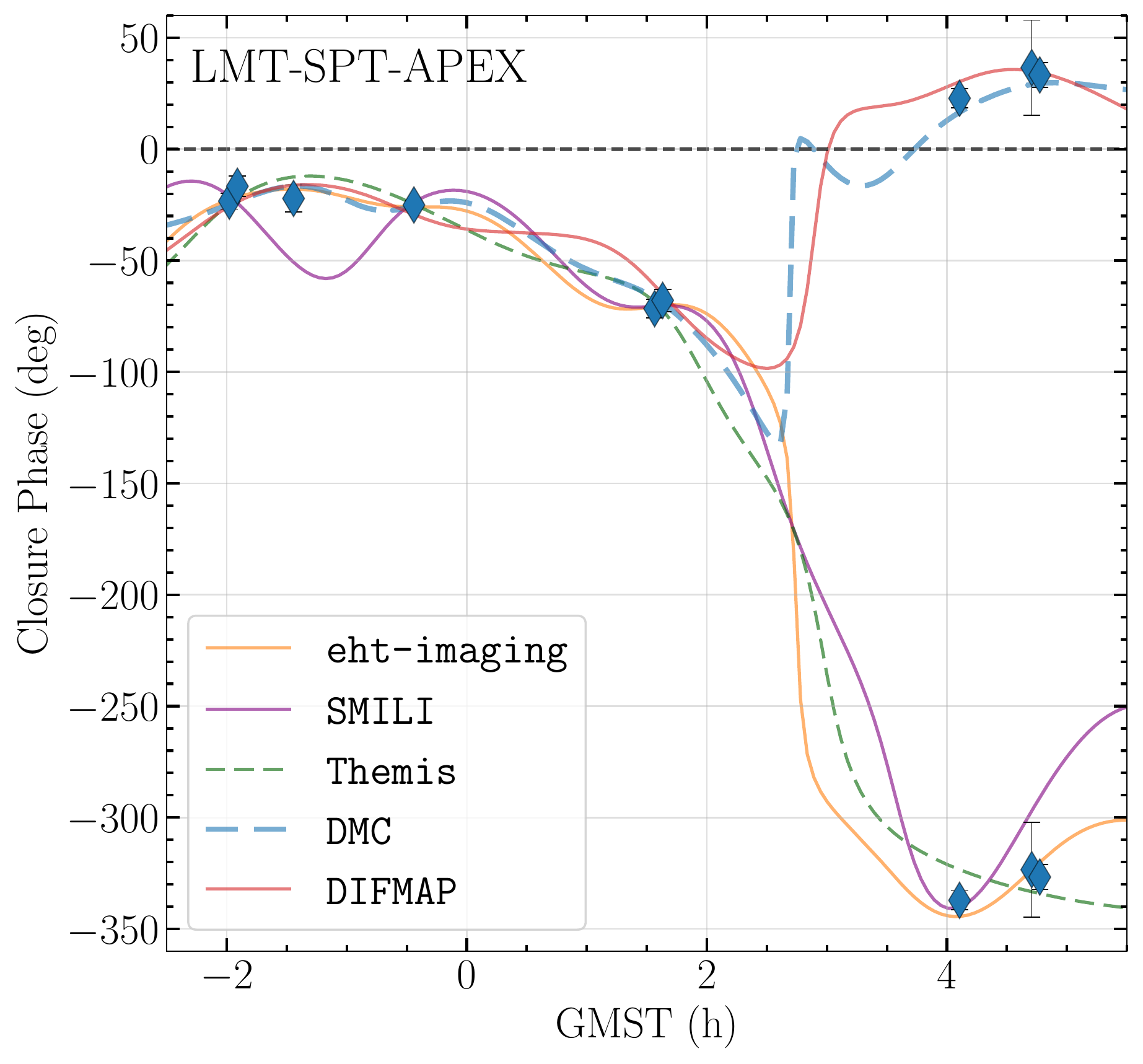}
    \caption{Comparison of observed closure phases on the LMT--SPT--APEX triangle on 2017 April 7 with predictions from the models shown in Figure \ref{fig:apr7_images}. The observations taken after GMST=3 are shown twice, with a 360 deg phase shift.}
    \label{fig:consistency_methods}
\end{figure}

\begin{figure*}
    \def\arraystretch{0.0}
    \setlength{\tabcolsep}{0.0pt}
    \begin{tabular}{ccc}
    \hspace{-0.0\linewidth}2017 April 6 & \hspace{-0.01\linewidth}2017 April 7 & \hspace{-0.08\linewidth}2017 April 11  \\
    \raisebox{0.13\linewidth}[0pt][0pt]{\rotatebox[origin=c]{90}{\tt eht-imaging}}\phantom{.} 
    \includegraphics[height=0.3\textwidth]{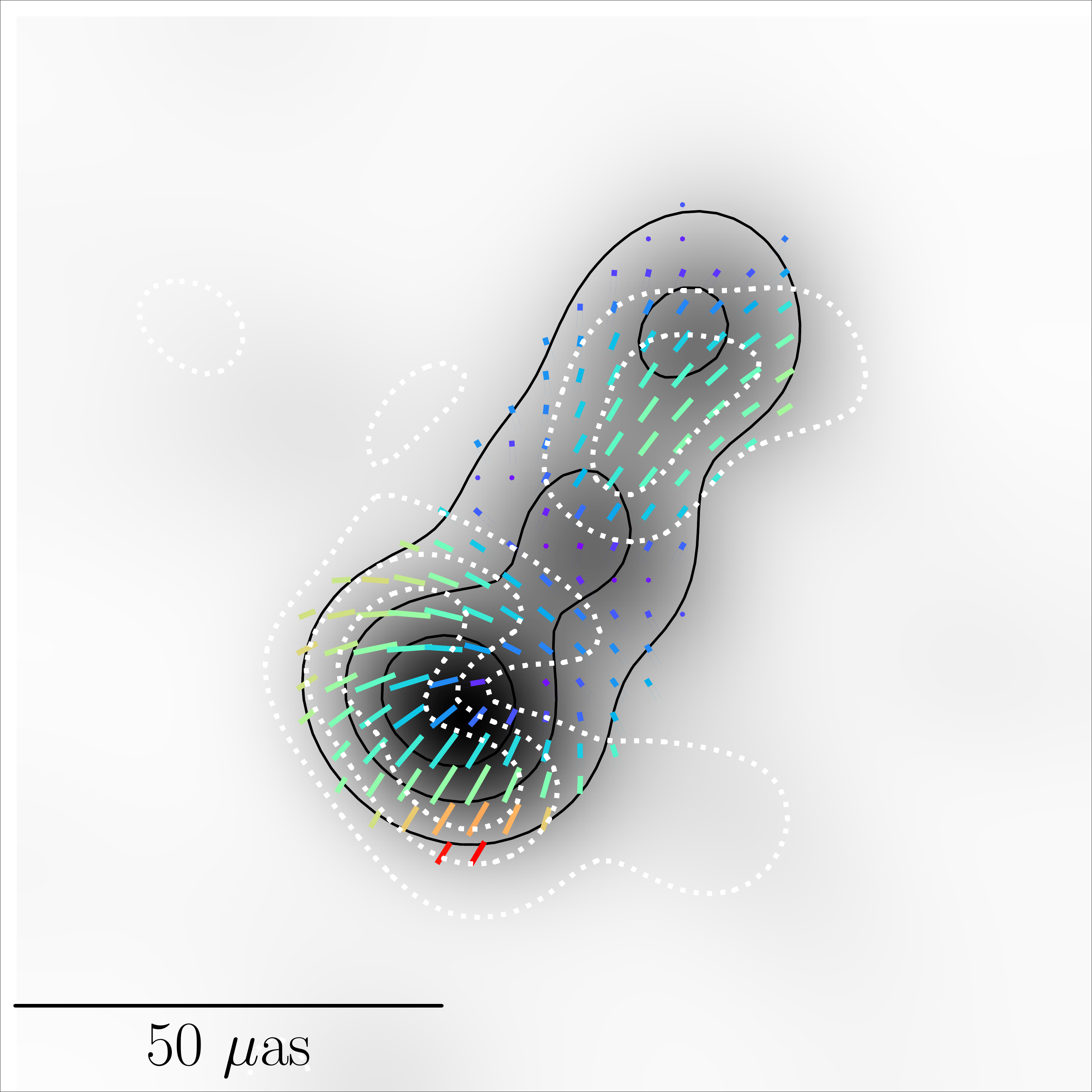}
    & \includegraphics[height=0.3\textwidth]{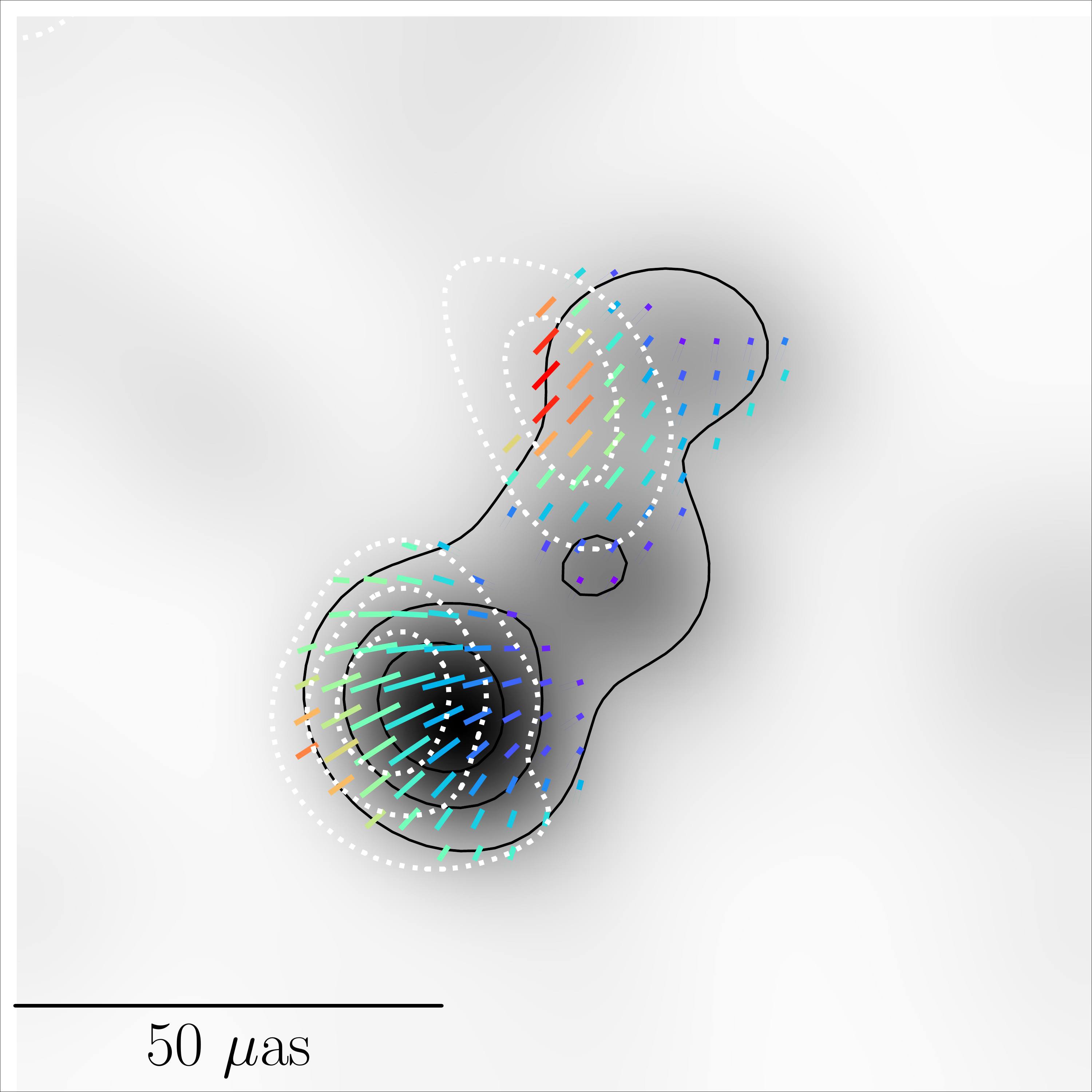}
    & \includegraphics[height=0.3\textwidth]{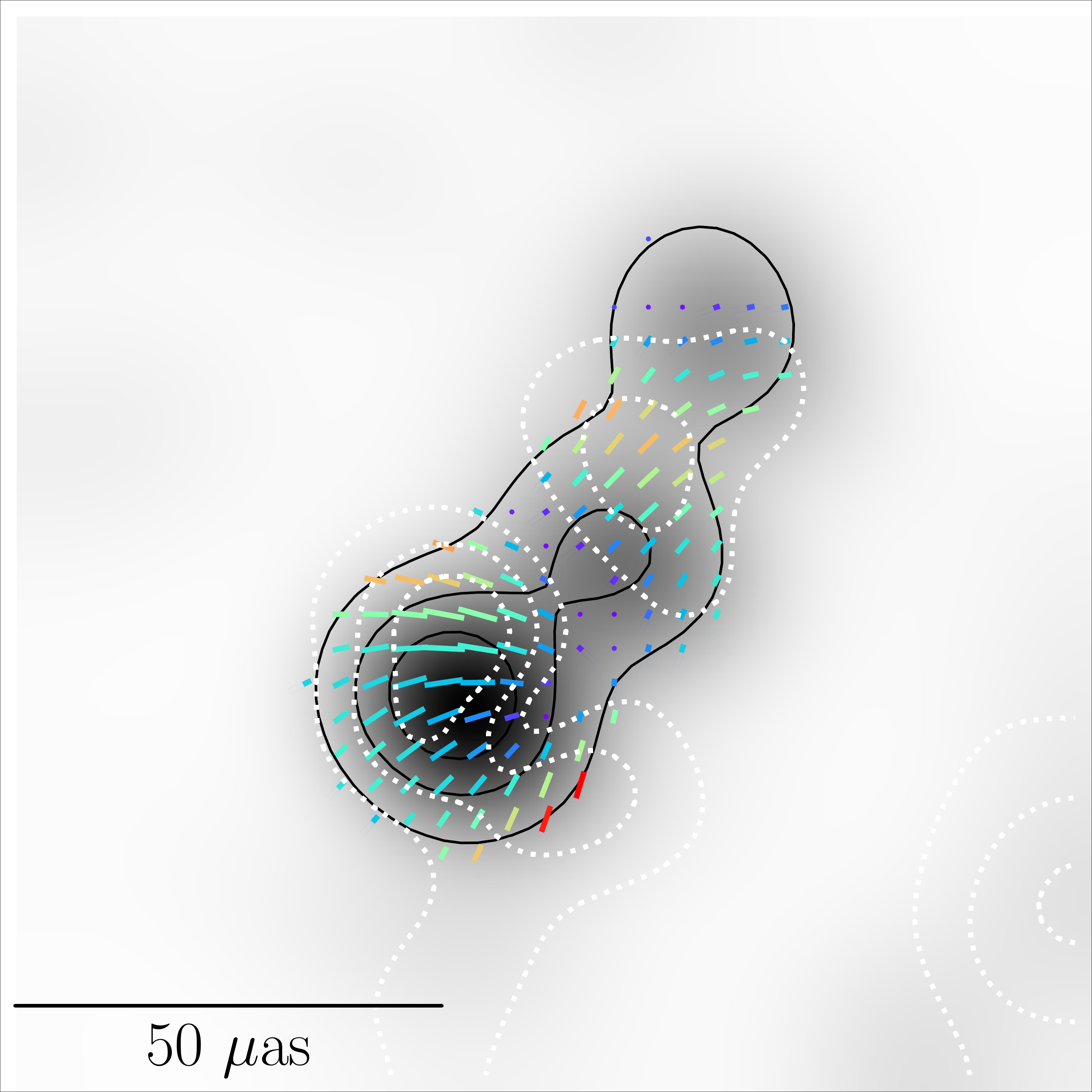}
    \includegraphics[height=0.3\textwidth]{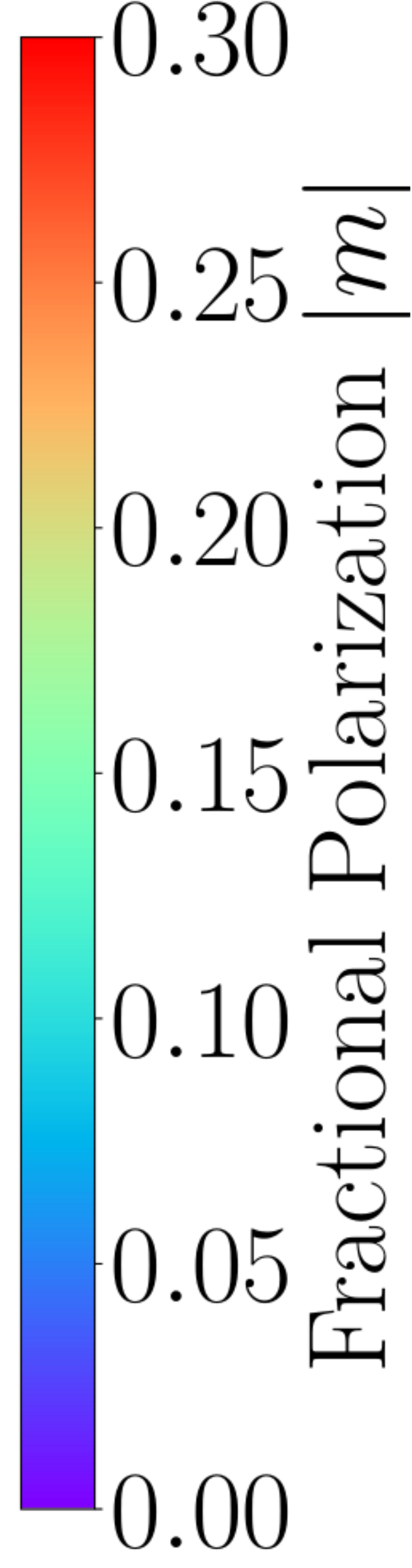} \\
    \raisebox{0.14\linewidth}[0pt][0pt]{\rotatebox[origin=c]{90}{\tt DMC}}\phantom{.}\hspace{0.1cm}
   \includegraphics[height=0.3\textwidth]{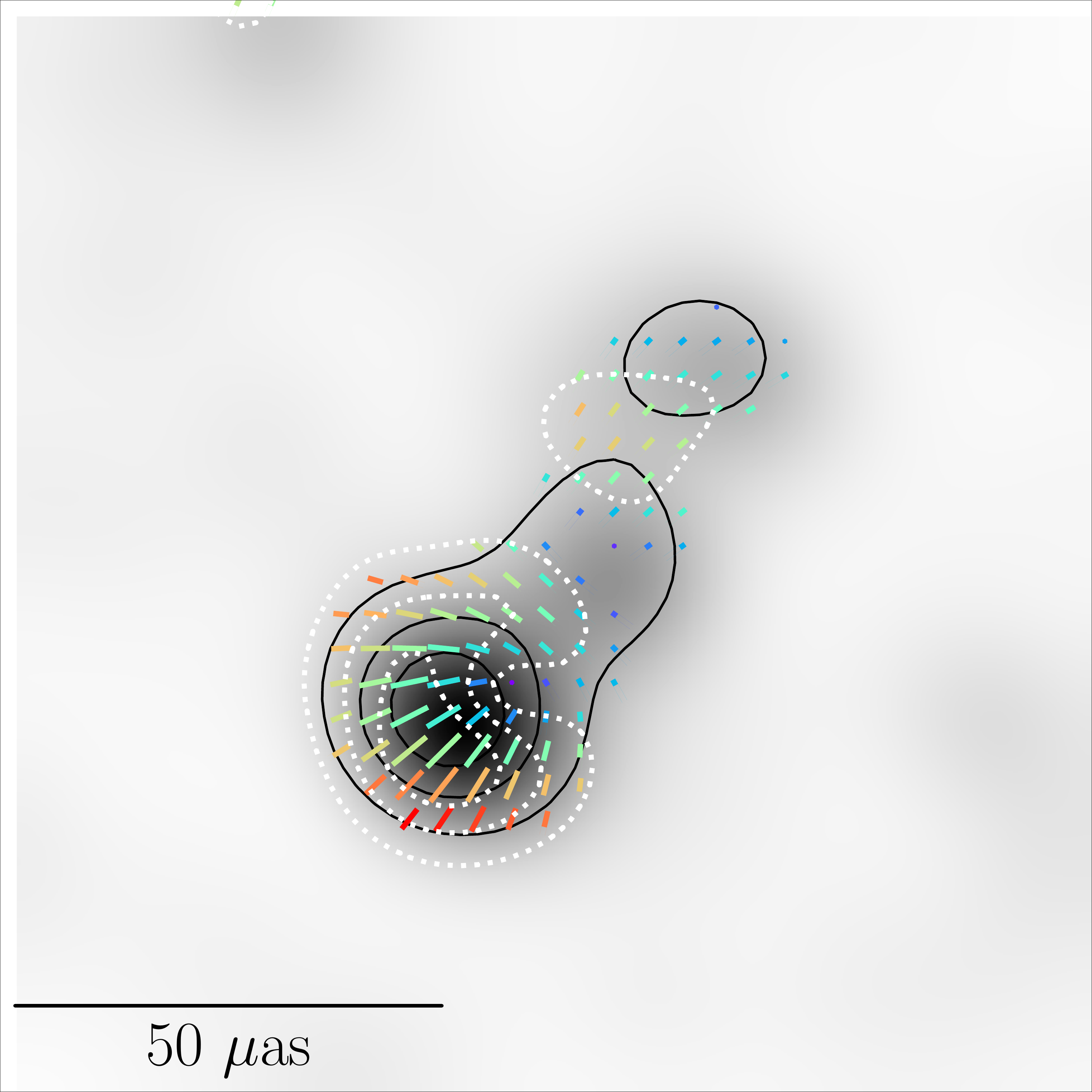}
   & \includegraphics[height=0.3\textwidth]{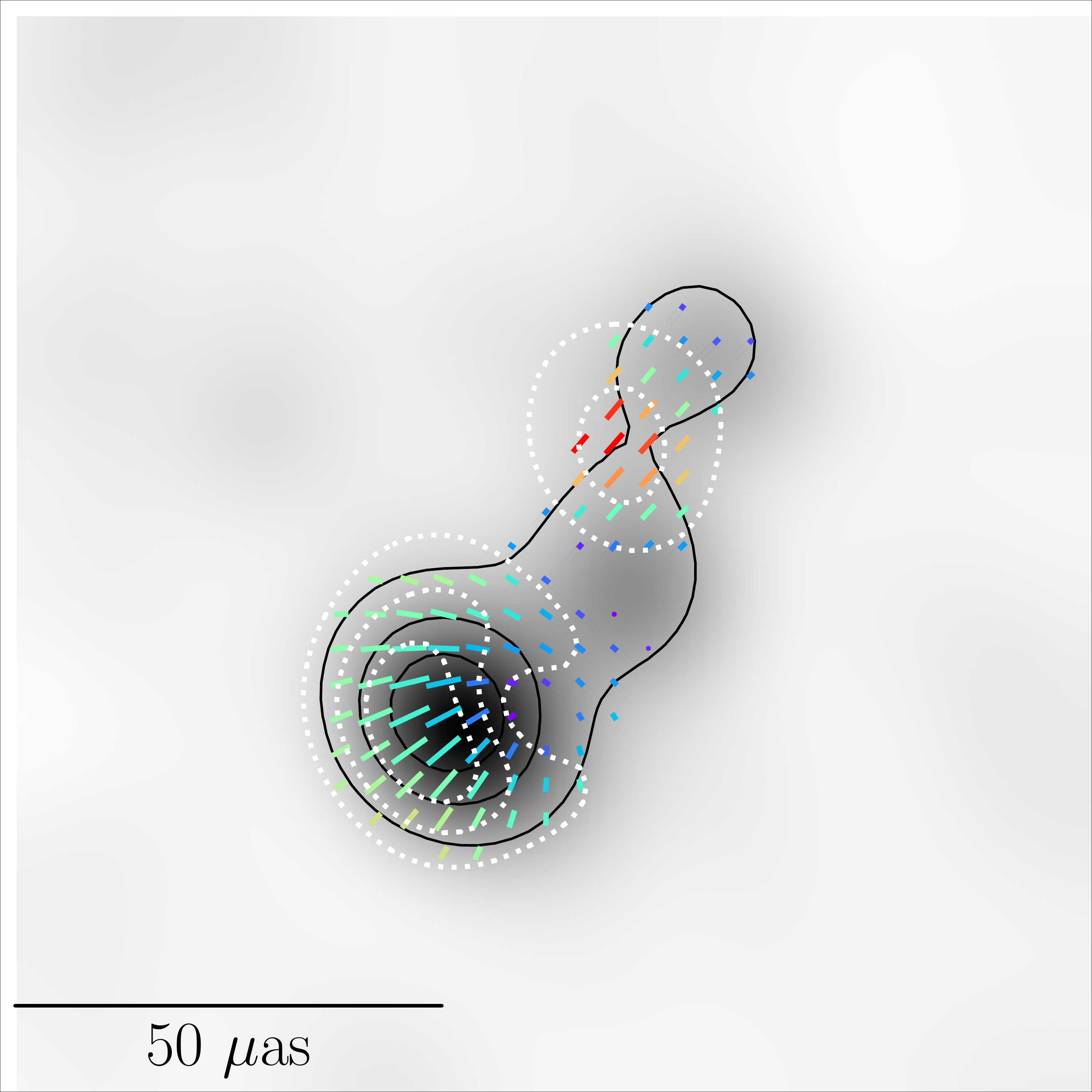}
   & \includegraphics[height=0.3\textwidth]{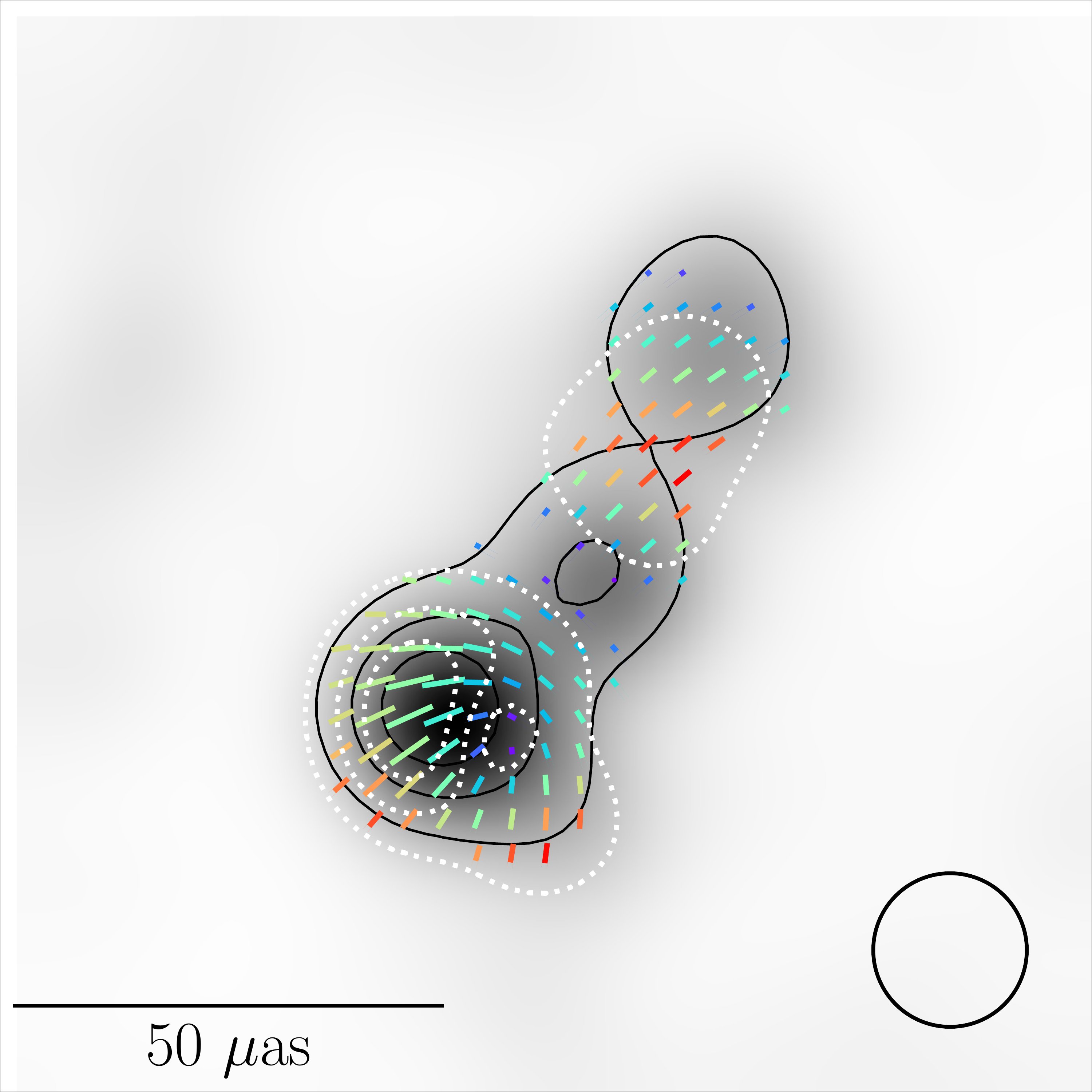}
    \includegraphics[height=0.3\textwidth]{pol_fraction_cbar.pdf}\\
    \end{tabular}
    \caption{Fiducial polarimetric images of \janine produced by the RML imaging method {\tt eht-imaging} and the posterior exploration method {\tt DMC}. The results are shown for the three \janine observation days that included ALMA. The total intensity is shown on a grayscale with black contours indicating 25, 50 and 75\% of the peak total intensity. White contours indicate 25, 50 and 75\% of the peak polarization intensity. The ticks show the orientation of the EVPA, their length indicates linear polarization intensity magnitude, and their color indicates fractional linear polarization. Cuts were made to omit all regions in the images where Stokes $\mathcal{I} <$ 10\% of the peak brightness and $\mathcal{P} <$20\% of the peak polarized brightness. The images are all displayed with a field of view of $128\,\mu$as, and all images are blurred with a circular Gaussian beam of $18\,\mu$as.
    }
    \label{fig:pol_images}
\end{figure*}


\section{EHT Image Analysis Results}
\label{sec:results}

\subsection{Total Intensity}
\label{sec:total_intensity}
The total-intensity analysis was performed on four individual observing epochs, 2017 April 5, 6, 7, and 10+11. The three imaging software packages used for the analysis of the EHT observations of M\,87$^*$ \citep{PaperIV} were employed to reconstruct total-intensity images of \janine: the Regularized Maximum Likelihood (RML) software {\tt eht-imaging} \citep{Chael_2016,Chael_2018} and {\tt SMILI} \citep{Akiyama_2017a,Akiyama_2017b}; and the CLEAN algorithm implemented in the {\tt DIFMAP} software package \citep{Shepherd_1997,Shepherd_2011}. Following the development of posterior exploration techniques for the M\,87$^*$ polarization results \citep{PaperVII}, we utilized two Markov chain Monte Carlo framework algorithms in addition to the imaging methods: {\tt DMC} \citep{Pesce2021} and {\tt Themis} \citep{broderick20,Broderick2020b}.  
The reconstructions typically combine both low and high band data sets, given the very small fractional bandwidth difference and high inter-band consistency reported in \citet{PaperIII} and \cite{PaperVII}.

The total-intensity structure of \janine at 230\,GHz is very resilient to various user-based choices in the imaging process, leading to an easily-recoverable three-component structure shown in Figures \ref{fig:apr7_images} and \ref{fig:avg_images}. While the reconstructed source morphology appears robust, the total compact flux density within the 200\,$\mu$as field of view and in the three components is more ambiguous. An upper limit is provided by the simultaneous connected-element interferometric-ALMA measurements reported in \citet{Goddi2021}, that is around 3.2\,Jy, with small day-to-day variations below 0.1\,Jy, within the calibration uncertainties. Different analysis pipelines recover anything between 2.0 and 3.2\,Jy within the 200\,$\mu$as field of view. Furthermore, the algorithms differ in their detailed approach to the image reconstruction. As an example, {\tt SMILI} favors image sparsity, keeping brightness in a compact sub-structures, while {\tt DMC} does not encourage sparsity in any way, possibly allowing for more flux density to be distributed throughout the field of view as a dynamic range-limited noise floor. Additionally RML methods typically assume compact imaging priors, discouraging emission further away from the core. For these reasons in Figure \ref{fig:apr7_images} {\tt SMILI} and {\tt DMC} have similar total brightness, but the main three-component structure appears significantly dimmer in the {\tt DMC} reconstruction. Furthermore, the total compact flux density ambiguity is a particularly severe problem for the EHT array, which for the observations of \janine has no baselines in the range between 2\,M$\lambda$ and 0.6\,G$\lambda$, a single SMT--LMT baseline in the range 0.6-1.5\,G$\lambda$ and no coverage between 1.5 and 2.8\,G$\lambda$. Hence, constraining structures larger than $\sim\,100\,\mu$as is extremely challenging with EHT data. 

Monitoring by the SMA\footnote{\url{http://sma1.sma.hawaii.edu/callist/callist.html?plot=1924-292}} shows that the total compact flux density at 1.3\,mm remained in the $3.2 \pm 1.0$\,Jy range since 2013 until the end of 2021, hence the EHT 2017 observations should correspond to a representative state of the source in this low luminosity period. In early 2009 the source went through a flaring phase, when the total flux density went up to about 10\,Jy. The proto-EHT VLBI results reported by \citet{Lu2012} correspond to this period. Interestingly, while there is much more flux density in the 2009 data set on short baselines (about 6\,Jy at 0.6\,G$\lambda$), the 2009 and 2017 data sets show a nearly consistent correlated flux of $\sim 1$\,Jy on the shared Hawai'i--SMT baseline (3-3.5\,G$\lambda$ in both epochs). This suggests that the 2009 flare event could be related to a more extended region, possibly further downstream from the 1.3\,mm VLBI core region.  

In Table~\ref{tab:chi-squares}, we show the reduced $\chi^2$ calculated for the low band data sets averaged in 120\,s bins, as a metric of the fit quality to closure quantities \citep[closure phases -- $\chi^2_{\rm CP}$, log closure amplitudes -- $\chi^2_{\rm LCA}$;][]{TMS, Blackburn2020} and visibility amplitudes ($\chi^2_{\rm AMP}$) for the final representative (or `fiducial') images from all methods and all observing days. The best reconstructions from the imaging methods based on least-squares fitting to closure quantities and the mean images from the posterior exploration methods were chosen as the fiducial images. The \texttt{eht-imaging} and \texttt{DMC} models exhibit the best values in the $\chi^2$ metric with consistently good performance for all days, and hence we focus on those two software packages in the subsequent quantitative analysis. In Figure~\ref{fig:apr7_images}, we show the fiducial images for all methods for the 2017 April 7 observations, restored to a resolution of 20\,$\mu$as. In Figure~\ref{fig:avg_images}, we show the method-averaged images across our four observing epochs (2017 April 10 and 11 are combined). The method-averaging procedure reduces the impact of method-specific systematics and provides a more conservative image-domain representation of the source, highlighting image features consistently reconstructed across different algorithms. However, the averaged image may fit the visibility domain observations to a lesser degree than the individual reconstructions, and hence more quantitative studies typically rely on the analysis of individual pipelines \citep[e.g.,][]{PaperIV,Kim2020,PaperVII}. The stability and robustness of the \janine image and derived amplitude gains were also confirmed in \citet{SgraP2} for the purpose of the calibrator gain transfer for the imaging of Sgr~A$^*$ \citep{SgraP3}. We verified that imaging merged data sets from separate days generally leads to a decrease in the fit quality, which justifies the choice to analyze observing days separately.

The total-intensity structure of \janine is very stable across the duration of the EHT 2017 campaign. Examples of this consistency are shown in Figure~\ref{fig:consistency}, where we plot closure phases \citep{TMS, Blackburn2020} from the EHT observations of \janine on two triangles along with the fits obtained by the \texttt{eht-imaging} pipeline for different observing days. In particular, the left panel of Figure~\ref{fig:consistency} presents closure phases on the sensitive ALMA--LMT--SMA triangle, showing good agreement of the models obtained for 2017 April 6, 7 and 11. The SMA--SMT--SPT closure phases (middle panel of Figure~\ref{fig:consistency}) are consistent in the first part of the track, but about GMST=4\,h the models diverge, with 2017 April 5 and 10-11 indicating a rapid closure phase growth, and 2017 April 6 and 7 a rapid decrease of the closure phase. After GMST=5\,h, the models are consistent again, as the phase is wrapping with a 360\,deg period. There is some evidence for structural evolution between 2017 April 5-7 and April 10-11, however, such phase degeneracies on triangles involving long baselines can be caused by small structural changes in the image domain, such as 
single $\mu$as-scale relative motion of the components \citep{Kim2020}. We quantify the image structure evolution in Section \ref{sec:parameter_extraction}. Furthermore, we can track down the degeneracy seen in the middle panel of Figure \ref{fig:consistency} to the absolute phase ambiguity seen on baselines between SPT and SMT/LMT/Hawai'i. In Figure \ref{fig:consistency_methods} we show that the same degeneracy can be seen on a single day between our five imaging pipelines, all fitting the available data well and resulting in very similar models, e.g., \texttt{eht-imaging} and \texttt{DMC} images seen in Figure \ref{fig:apr7_images}.

\subsection{Linear polarization}
\label{sec:linear_polarization}
Due to the high consistency between all five methods in total intensity imaging,  we employed one RML imaging method ({\tt eht-imaging}) and one posterior exploration method ({\tt DMC}) exhibiting particularly good values of $\chi^2$ (see Table \ref{tab:chi-squares}) to streamline the polarimetric imaging and analysis. Both software packages have been extensively tested, including their performance on polarimetric reconstructions of synthetic images, see Appendix J of \citet{PaperVII}. We focus on the three observing days which include the sensitive ALMA array (2017 April 6, 7, and 11). Additionally, the presence of ALMA in the array enables a straightforward calibration of the absolute electric vector position angle (EVPA). In the right panel of Figure~\ref{fig:consistency} we demonstrate the consistency of the RML models with the polarimetric data. We are using the absolute value of the Fourier domain fractional polarization $\breve{m}$ \citep{Johnson_2015},
\begin{equation}
    \breve{m} = \frac{\hat{\mathcal{Q}} + i\,\hat{\mathcal{U}}}{\hat{\mathcal{I}}} \ 
\label{eq:mbreve}
\end{equation}
on the very sensitive ALMA--LMT baseline, where $\hat{\mathcal{I}}$, $\hat{\mathcal{Q}}$, and $\hat{\mathcal{U}}$ are Fourier domain Stokes components of the radiation field. Given the rotation measure value of $\sim 4 \times 10^4$\,rad m$^{-2}$ reported in \citet{Goddi2021}, we expect that Faraday rotation effects do not affect the observed EVPA by more than 5 deg at the observing frequency of 230\,GHz.

\begin{figure}
    \centering
    \includegraphics[width=1.0\linewidth]{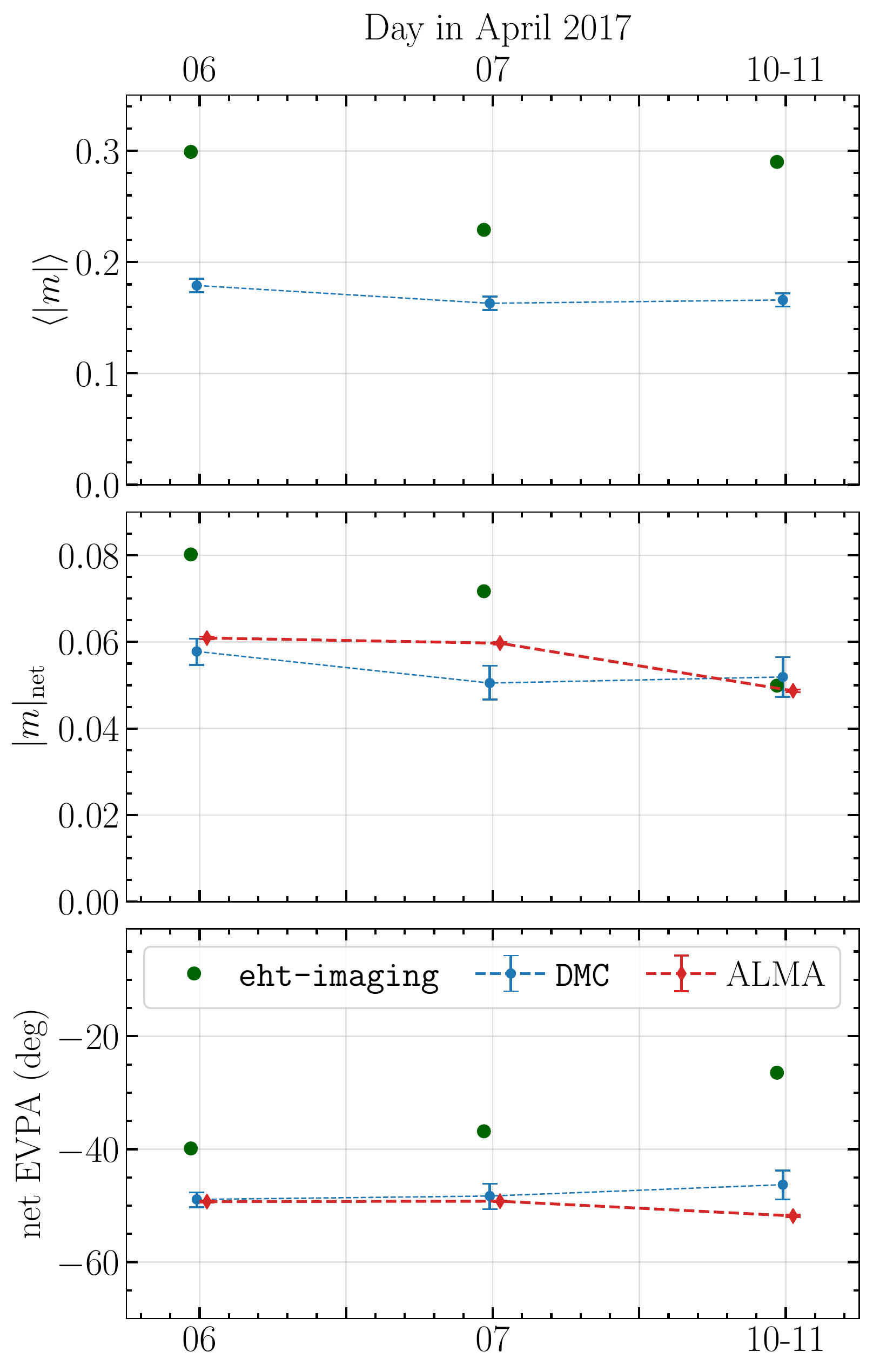}
    \caption{Image-averaged linear polarimetric properties of the source. Blue points with errorbars represent 68\% confidence intervals from the {\tt DMC} posterior distributions. Green points without errorbars correspond to a single measurement from the {\tt eht-imaging} images. Red points correspond to ALMA measurements from \citet{Goddi2021}. For the sake of clarity a small horizontal shift has been added between the markers representing different methods within the same day.
    }
    \label{fig:total_pol}
\end{figure}

In Figure~\ref{fig:pol_images}, we show the fiducial polarimetric images of \janine for the three days with ALMA and two analysis pipelines. The linearly polarized emission is localized in the VLBI core (the brightest and southernmost component $C0$, see Section~\ref{sec:parameter_extraction}) and in-between the second and third total-intensity components along the jet direction. In these two regions, the resolved image-domain fractional polarization reaches $\sim20$\%.
While the EVPAs seen in the outer jet components are mainly aligned parallel to the jet axis, the EVPA pattern in the core region rotates in a fan-like pattern, causing depolarization in the image-integrated results.
These features are persistent for all observing days and across methods.

We can characterize the image-integrated  linear polarization (over a 200\,$\mu$as field of view) with the following metrics. First, the intensity-weighted average polarization fraction across the resolved EHT image (we blur models with a 15\,$\mu$as circular Gaussian beam) is given by
\begin{equation}
    \langle|m|\rangle = \frac{\sum_{k} \sqrt{\mathcal{Q}_k^2+\mathcal{U}_k^2}}{\sum_k \mathcal{I}_k}
    = \frac{\sum_{k} \mathcal{P}_k}{\sum_k \mathcal{I}_k}
    \ .
    \label{eq:mavg}
\end{equation}
Here $\mathcal{I}$, $\mathcal{Q}$ and $\mathcal{U}$ are the image-domain Stokes parameters, and the sums are taken over all pixels in the image. In the first panel of Figure \ref{fig:total_pol} we show $\langle|m|\rangle$ across observing days and imaging pipelines to be about 20-30\,\%. In Figure \ref{fig:pol_images} we see that the largest contribution to $\langle|m|\rangle$ comes from the core region.

A second metric is the coherently-averaged polarization fraction $m_\mathrm{net}$, representing the unresolved fractional polarization in the 200\,$\mu$as field of view,
\begin{equation}
    m_\mathrm{net} = \frac{\sum_k \mathcal{Q}_k + i \sum_k \mathcal{U}_k}{\sum_k \mathcal{I}_k} \ .
    \label{eq:mnet}
\end{equation}
In the middle and bottom panels of Figure \ref{fig:total_pol} we show the absolute values $|m|_{\mathrm{net}}$ and EVPAs of $m_\mathrm{net}$ at about 6\% and -50 deg, respectively. Comparing these results with the ALMA measurements reported by \citet{Goddi2021} we find overall good consistency, in particular with the {\tt DMC} pipeline. This indicates that the polarized emission unresolved by ALMA with $\sim$\,arcsec resolution is mostly confined to the narrow field of view of the EHT. The large difference between $|m|_{\mathrm{net}}$  and $\langle|m|\rangle$ is related to the large EVPA variation in the image, as seen in the core component in Figure \ref{fig:pol_images}.

\subsection{Circular polarization}
\label{sec:circular_polarization}
As \texttt{DMC} is a posterior-exploration code performing full-Stokes modeling \citep{Pesce2021}, we can use the recovered posterior distributions to determine whether the 230\,GHz images contain statistically significant detections of circular polarization. Moreover, in contrast to other analysis methods that we consider, \texttt{DMC} explores relative $R/L$ complex polarimetric station gains as parameters of the fitted image, thus providing more robustness against systematic uncertainties in the polarimetric a priori flux density calibration.
We measure the detection confidence in the resolved image as the ratio between the local mean posterior and the local posterior standard deviation of the estimated circular polarization, evaluated based on 1000 images drawn from the posterior distribution. Our confidence does not exceed $2\,\sigma$ at any location in the image on any observing day. The distribution of net (image-averaged) circular polarization is also consistent with zero, in agreement with the findings of \citet{Goddi2021}. For comparison, the same procedure determines the confidence in the source structure corresponding to over 30\,$\sigma$ when applied to Stokes $\mathcal{I}$ images, and about $10\,\sigma$ when applied to linear polarization images. 

\begin{figure}
    \centering
    \includegraphics[width=\linewidth]{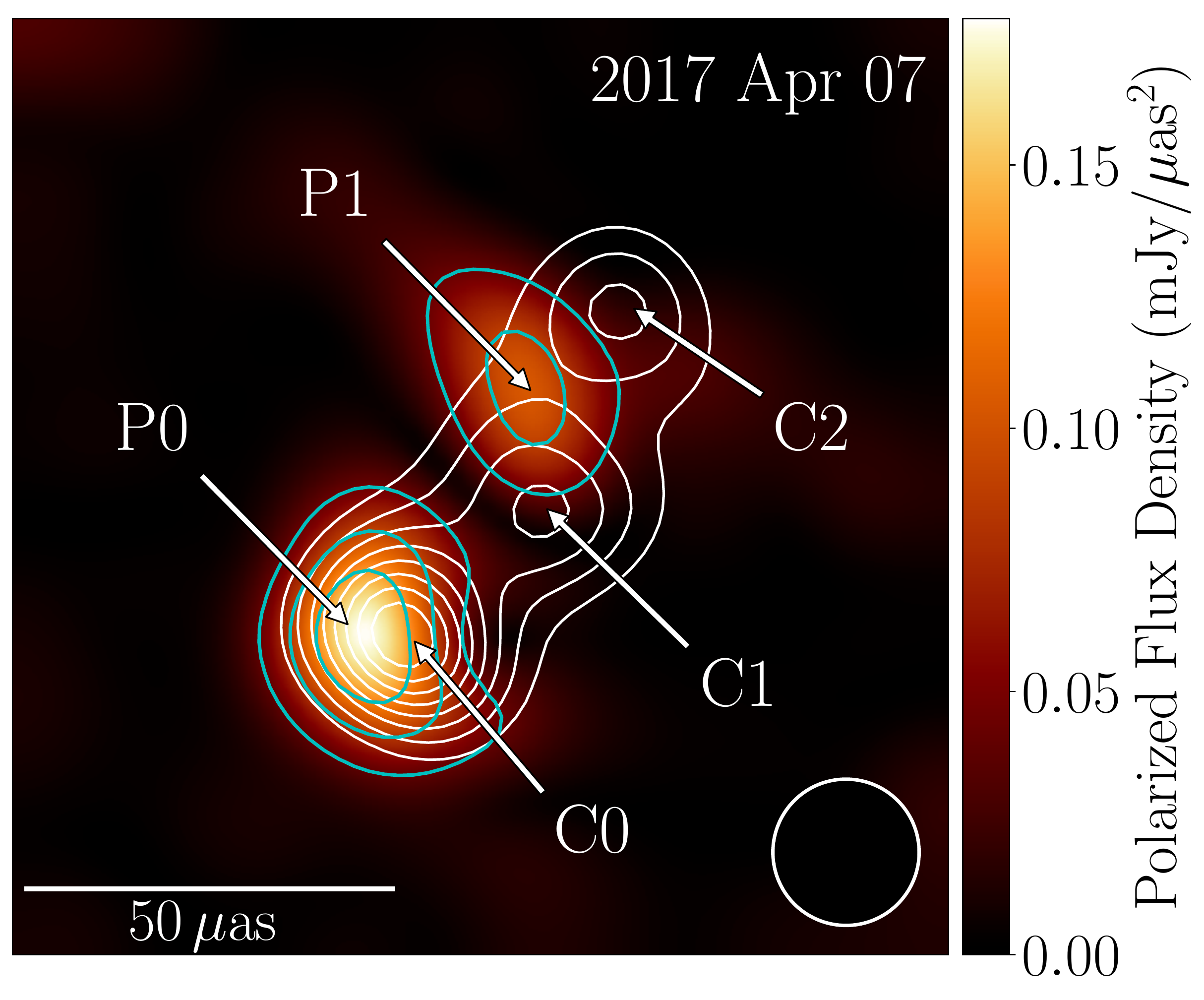}
    \caption{Schematic of the total-intensity ($C0$, $C1$, and $C2$) and linear polarization ($P0$ and $P1$) components in the EHT images of J1924--2914 at 230\,GHz. We show the method-average total intensity and linear polarization image of our best observation day, 2017 April 7. Linear polarization contours are shown in cyan, denoting regions of 25, 50 and 75\% of the peak linearly polarized intensity. Guiding contours in total intensity are shown in white, denoting regions from 20\% to 90\% of the peak total intensity, in steps of 10\%. The nominal EHT beam is shown as a white circle in the lower right.}
    \label{fig:components}
\end{figure}
\subsection{Image components feature extraction}
\label{sec:parameter_extraction}

In Figure~\ref{fig:components} we show a diagram of the different components in total intensity and linear polarization on which image domain-based feature extraction was performed across observing days and methods. 
We identify the brightest total intensity component at the south-eastern end of the jet as the 
VLBI core $C0$, following the image morphology at lower frequencies, and the other total intensity features as
the first jet component $C1$, and the second jet component $C2$. The components showing the highest
linear polarization are $P0$ in the core region, and the jet component $P1$, the latter located between $C1$ and $C2$. 

\begin{figure}
    \centering
    \includegraphics[width=1.01\linewidth]{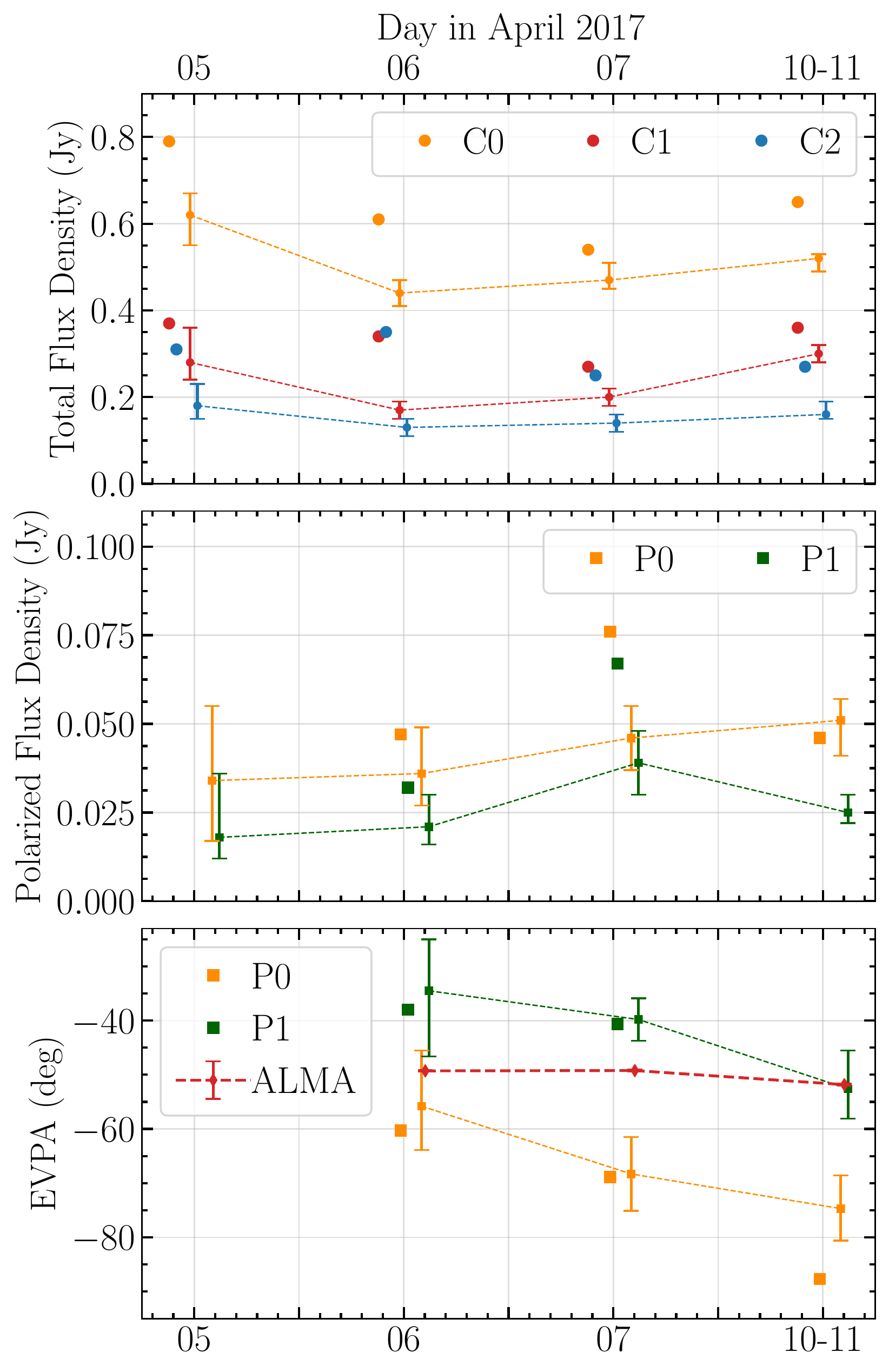}
    \caption{Total intensity and linear polarization properties of the individual components identified in Figure~\ref{fig:components}. The different colors correspond to the different total intensity (circles) and linear polarization (squares) components. Points with errorbars represent 68\% confidence intervals from the {\tt DMC} posterior distributions. Points without errorbars correspond to a single measurement from the {\tt eht-imaging} images. For the sake of clarity a small horizontal shift has been added between the markers representing different methods and components within the same day. 
    }
    \label{fig:featuresA}
\end{figure}

\begin{figure}
    \centering
    \includegraphics[width=0.96\linewidth]{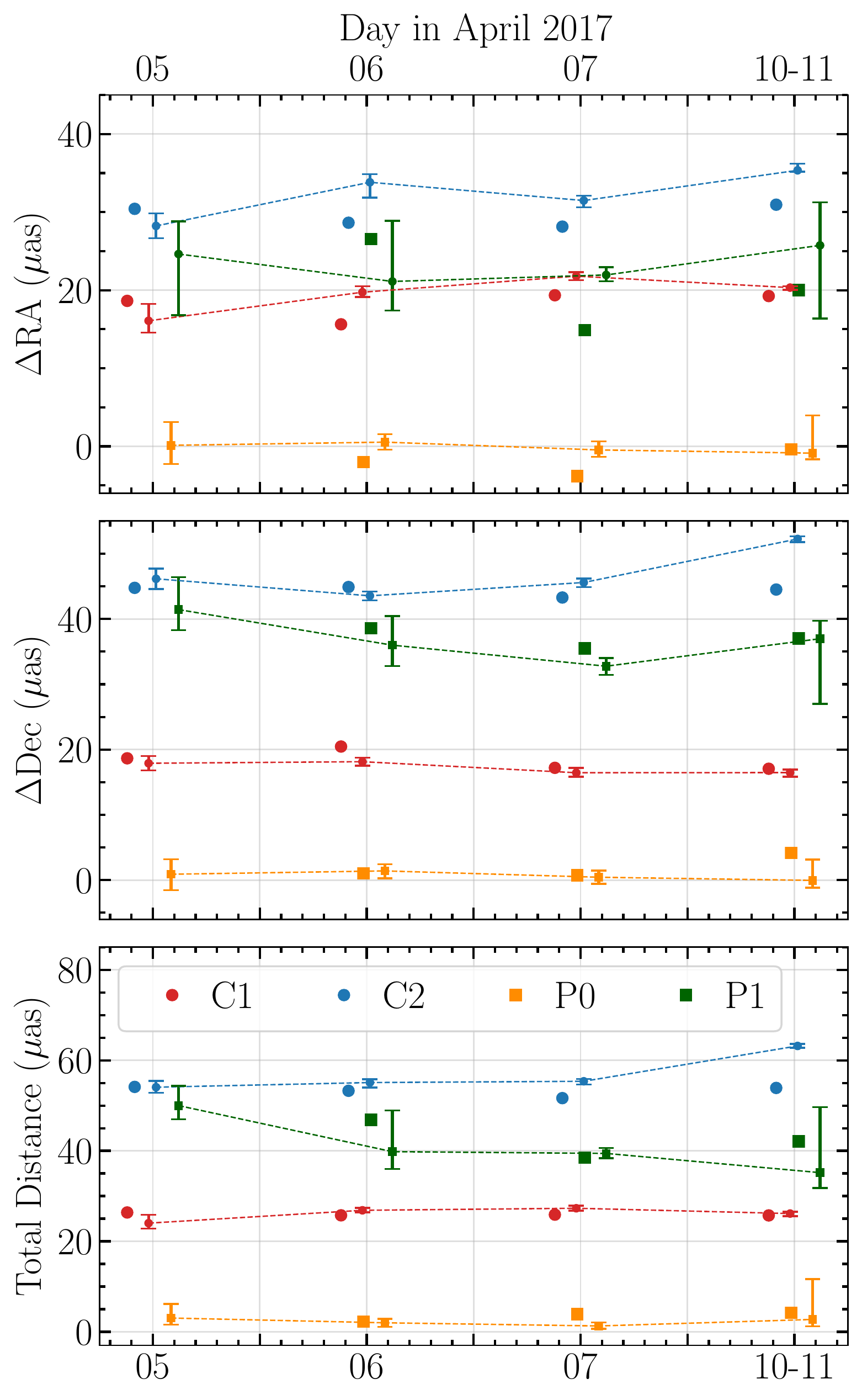}
    \caption{Relative distance of the total-intensity ($C1, C2$) and linear polarization ($P0, P1$) components from the VLBI core $C0$ across the EHT observing campaign. The component labels follow that of the schematic in Figure~\ref{fig:components}.
     For the sake of clarity a small horizontal shift has been added between the markers representing different methods and components within the same day.
    }
    \label{fig:featuresB}
\end{figure}

In case of {\tt DMC}, similarly as in the Section \ref{sec:circular_polarization}, we consider 1000 images drawn from the posterior distribution. For each image, we perform an image-domain brightness maxima search, identifying them with the maxima of components indicated in Figure~\ref{fig:components}. We then define binary masks around each extremum to compute the total flux density corresponding to each component (Figure \ref{fig:featuresA}), and the positions of the flux density centroids (Figure \ref{fig:featuresB}). The errorbars reflect the 68\% confidence intervals of individual measured quantities extracted from {\tt DMC} posteriors. {\tt DMC} enables parameter extraction for all epochs, apart from EVPA measurements on April 5, which is the day without ALMA participation and hence lacks the absolute EVPA calibration. Additionally we show a similar measurement obtained from a single {\tt eht-imaging} reconstruction per day (markers without the errorbars in Figures \ref{fig:featuresA}-\ref{fig:featuresB}). The total intensity {\tt eht-imaging} results are available for all epochs, while polarization results are only available for the days with ALMA.

\begin{figure*}[t]
    \centering
    \raisebox{-0.3em}[0pt][0pt]{\includegraphics[width=0.2695\textwidth]{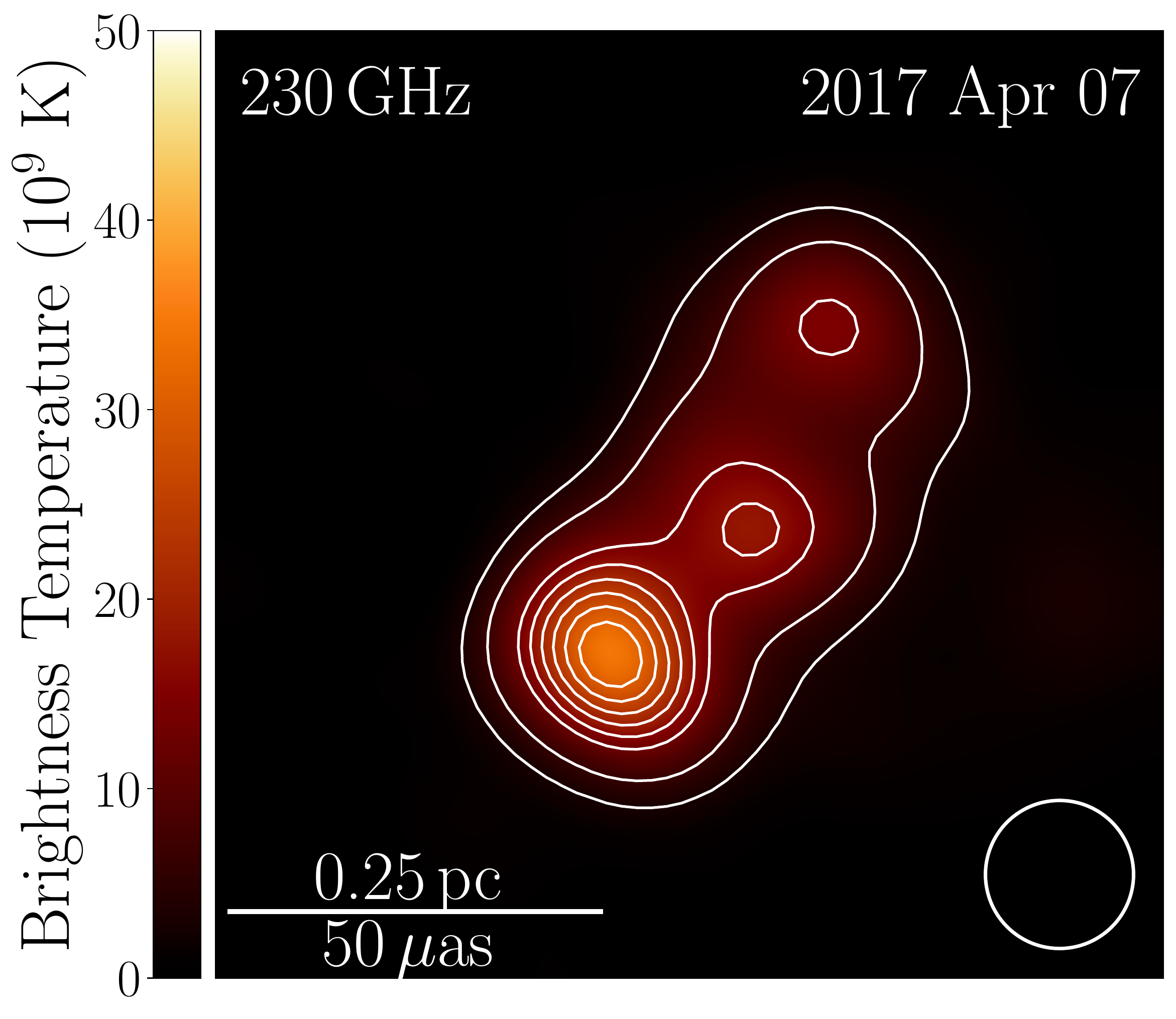}}
    \includegraphics[width=0.224\textwidth]{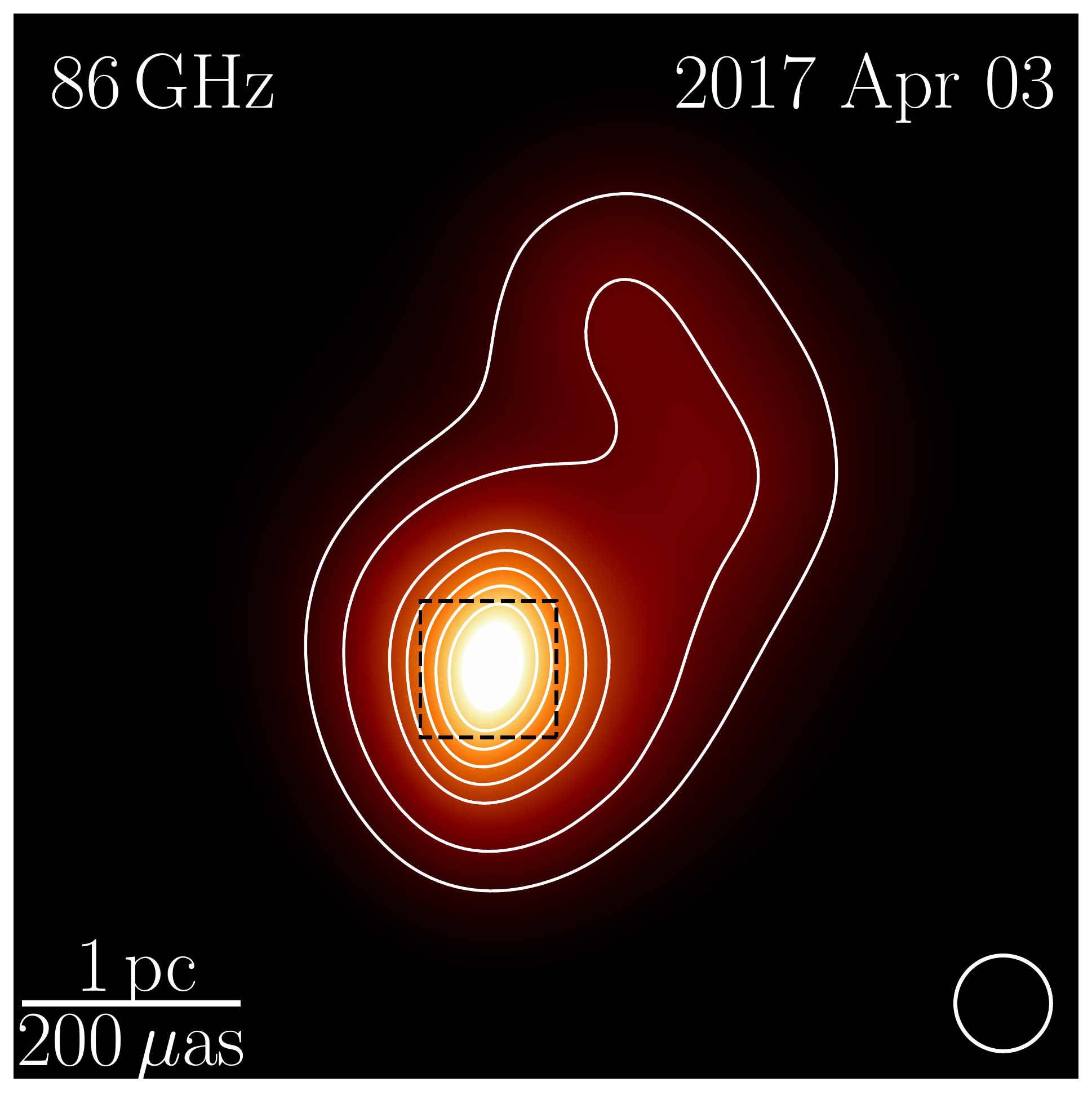}
    \includegraphics[width=0.223\textwidth]{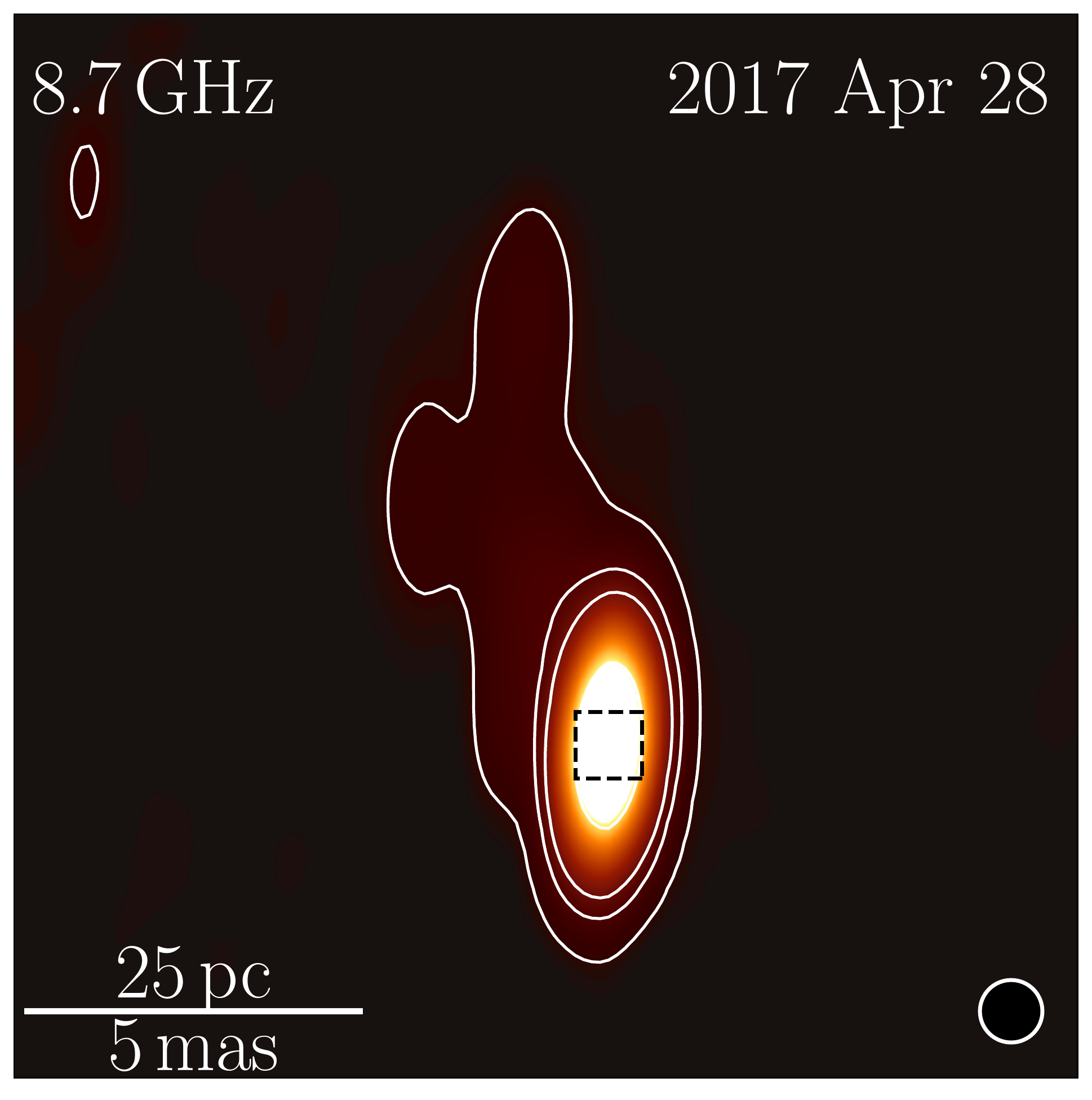}
    \raisebox{-0.3em}[0pt][0pt]{\includegraphics[width=0.2695\textwidth]{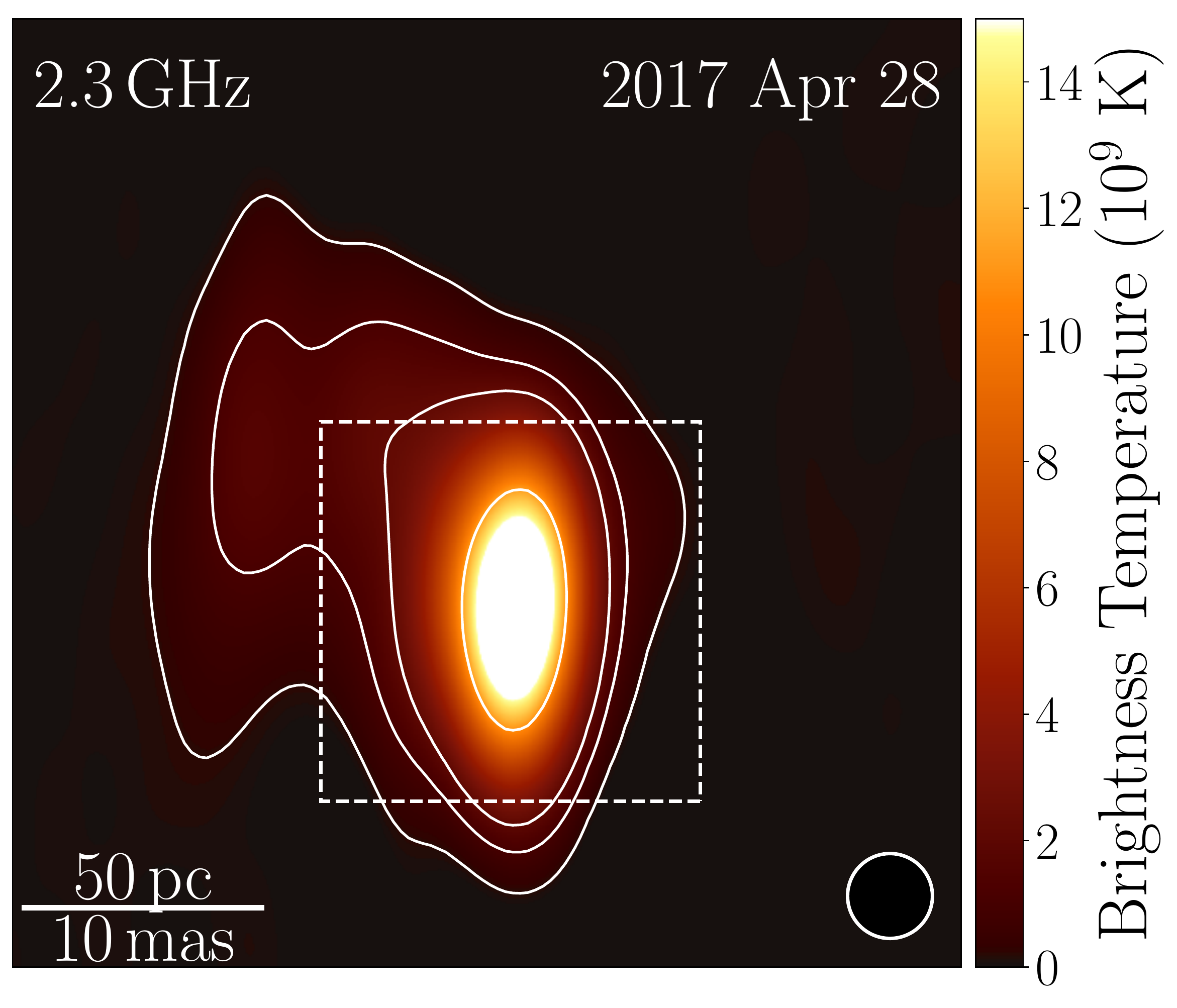}}
    \caption{Multi-frequency images of J1924--2914 from the 2017 April observations. From left to right: the 230\,GHz 2017 April 7 EHT image; the 86\,GHz 2017 April 3 GMVA+ALMA image; and 8.7 and 2.3\,GHz images taken on 2017 April 28 with the VLBA. The contours delimit regions from 10\% to 90\% of the peak brightness temperature, in steps of 10\%. To make all the features visible, the EHT image and lower-frequency images have different intensity ranges. The EHT image is that of Figure~\ref{fig:avg_images} and follows the brightness temperature range on the left, whereas the three images at lower frequencies have the brightness temperature range of the color bar on the right. The white circles in the lower left of each panel denote the nominal instrument resolution. The dashed squares delimit the field of view of the preceding higher frequency image. }
    \label{fig:multifreq}
\end{figure*}

In Figure~\ref{fig:featuresA}, we present the total-intensity and polarization properties of the components identified in Figure~\ref{fig:components} across the four observing days. We show the variation of total flux density, linear polarization flux density and EVPA integrated over each component across the EHT observing campaign extracted from the {\tt eht-imaging} and {\tt DMC} imaging results. The southernmost component $C0$ has the highest total flux density and is assumed to be the 230\,GHz VLBI core. For the component-integrated EVPA, we also compare the results with the image-integrated EVPA measured from simultaneous interferometric-ALMA observations in \citet{Goddi2021}. We notice a bias between component flux densities identified with \texttt{eht-imaging} and \texttt{DMC}, with the former returning values larger by about 25\% for $C0$ and $C1$ components. We attribute the effect to imaging algorithm systematics, particularly to sparsity-based regularization employed for the \texttt{eht-imaging} reconstructions, see the discussion in Section \ref{sec:total_intensity}. Given the associated uncertainties, there are no strong indications of the time evolution of the component flux density on the timescale of our EHT observations. The core component $C0$ is about 2 times brighter than $C1$ and $C2$, the latter two showing similar flux densities. While $P0$ has a higher polarized flux density than $P1$, the EVPA varies by about 90 deg across the compact $P0$ core component, consistently between days and methods (see Figure~\ref{fig:pol_images}), adding destructively in Figure~\ref{fig:featuresA}. When we add absolute values of the linear polarization instead of coherently adding complex numbers, we find about 0.1\,Jy in the $P0$ region for images blurred to a 15\,$\mu$as resolution. There is also a notable EVPA rotation trend observed in both $P0$ and $P1$ components and not observed in the ALMA-only data (bottom panel of Figure~\ref{fig:featuresA}). 
While this feature is not very statistically significant and may be a statistical fluke, it may also indicate some systematic VLBI calibration bias. However, the bias in the absolute EVPA calibration, which follows the ALMA QA2 calibration \citep{Goddi2021}, would be expected to impact all EHT sources in a similar fashion, and an opposite trend in EVPA was found for M\,87$^*$ \citep{PaperVII}. Since the net EVPA of VLBI images is consistent with the ALMA-only measurements (at least for the DMC images, see Figure \ref{fig:total_pol}), this implies that the EVPA shift of the components is compensated by the change in the residual net EVPA within the field of view. Hence, if the effect is systematic, it appears to be related to the imaging or parameter extraction algorithms, rather than to the data calibration. In any case, we conclude that a $< 20$\,deg consistency between the ALMA-only data and the sum of the VLBI components is overall reassuring and constrains the contribution from the systematic errors.

In Figure~\ref{fig:featuresB}, we present the relative distance of the individual total intensity and polarization component centroids from the core $C0$ across the EHT observing campaign. There is no systematic offset between centroids of $C0$ and $P0$, indicating that most likely they correspond to the same physical core component. However, the peak brightness location of $P0$ is shifted to the East with respect to the peak brightness of $C0$, see also the discussion in Section \ref{sec:core}. We see no significant motion of the components between 2017 April 5 and 11, particularly $C1$ is well constrained and on all days and for both pipelines its distance from the core remains consistent within 2\,$\mu$as. For \janine, observed motion of 1\,$\mu$as/day corresponds to an apparent velocity of 8\,$c$, which translates to an upper limit on apparent velocity $\beta_{\rm app}(C1) < 2.7\,c$. Kinematic analysis at 15\,GHz \citep{Lister2019} resulted in apparent velocities of three features, yielding $\beta = 7\,c$ at a distance of 5\,mas from the core, and two features on sub-mas scale, moving with $\beta = 2.6\,c$ and $\beta = 0.2\,c$. While a direct comparison of the apparent velocities seen at 15\,GHz and 230\,GHz is difficult in view of the bent jet morphology and different regions being probed, let us engage in some plausible speculation about this. The wider range of apparent velocity estimated in the innermost region may suggest jet bending and a smaller jet inclination of the innermost region (bending away from the line of sight). In this regime, $\beta_{\rm app} \approx \theta/(1-\beta)$ and thus small intrinsic variations of $\theta$ can result in relatively large variations of the apparent velocity. At present, the physical origin of the relative large variation of the component speeds along the jet is unclear. It could be due to component motion along spatially bent trajectories, but intrinsic jet acceleration combined with regions of slower velocity or even stationarity (shocks) also cannot be excluded. 
Future more detailed kinematic studies will be required to clarify this.

We also note a peculiarity seen in Figure~\ref{fig:featuresB}, resulting mostly from the \texttt{DMC} analysis: there is marginal evidence that, while component $C2$ separates from the core and moves downstream, the motion of the $P1$ centroid goes in the opposite direction, approaching the VLBI core. Whether this is due to pattern motion or is a projection effect in a rotating jet remains at this time an open question. We can only place very loose upper limits on the apparent velocities with respect to $C0$, $\beta_{\rm app}(C2) < 13\,c$, $\beta_{\rm app}(P1) < 27\,c$ with a hint of acceleration, which may be related to the transverse shock discussed in Section \ref{sec:jet_features}, rather than to the overall acceleration pattern expected in the inner part of the ejected jet.

\section{Discussion}\label{sec:discussion}

\subsection{Multifrequency images of the bent jet}\label{sec:multifreq}

In Figure~\ref{fig:multifreq}, we show multi-frequency images of \janine from close in time observations in April 2017 with the VLBA (2.3 and 8.7\,GHz), the GMVA+ALMA (86\,GHz), and the EHT (230\,GHz). Ranging two orders of magnitude in frequency, these images show jet structure spanning from sub-parsec to 100 parsec scales. The EHT image of \janine provides an unprecedented view of the inner parsec of this blazar. 

The projected position angle (PA) of the jet gradually rotates counter-clockwise with increasing distance from the jet base. In the 2.3\,GHz VLBA image the PA is $50\pm5$\,deg east of north at about 20\,mas from the core and the jet bends toward the north-south direction closer to the core, consistently with the jet orientation in the non-simultaneous 1.6\,GHz image at 5-10\,mas from the core \citep{Shen_1999}.
At 8.7\,GHz we find a PA of $25 \pm 5$\,deg at about 5\,mas from the core, consistent with the archival VLBA monitoring results at 15\,GHz, showing a persistent jet orientation at a PA of about 30\,deg in the epochs from 1995 to 2013 at angular scales $\sim\,5\,$mas \citep{Kellermann1998,Pushkarev2017}. The same MOJAVE observations hint at more variability of the jet PA on the smallest resolved scales $\sim\,1\,$mas. Observations by \citet{Shen_2002} with the VLBA between 1994 and 2000 across four frequencies (5, 12, 15, and 43\,GHz) showed a consistent PA orientation of 30\,deg at 5\,GHz with a clockwise shift of about 51-67\,deg at 43\,GHz. At 86\,GHz a possible bent jet structure is seen, with the inner jet oriented with a PA of about $-40$\,deg less than $0.3$\,mas from the core, and an apparent 
transition to a north-east direction further out. The $-40$\,deg PA with respect to the core is consistent with the PA of the single jet component located $\sim 400\,\mu$as from the core, imaged from 2018 April GMVA observations, see Figure 4 of \citet{Issaoun2021}. 2018 images at 86\,GHz do not indicate jet bending. At 230\,GHz, the component $C1$ is located at a PA of $-45$\,deg and $C2$ at a PA of $-35$\,deg with respect to the core $C0$. This morphology can potentially be explained by a helical structure in the jet \citep{Conway1993,Steffen1995}. Such a helical jet structure can be caused by an orbiting lower-mass secondary black hole around a stable primary central black hole, as also proposed for 4C\,73.18 \citep{Roos1993} and OJ\,287 \citep[e.g., ][]{Dey_2021,Gomez_2021}, precession caused by a wobbling disk \citep{Britzen2018}, or a large-scale accretion flow that is tilted with respect to the black hole spin axis \citep[e.g., as in M\,81;][]{Marti_2011}.

Alternatively, the jet could be showing a sharp bend related to a collision between the jet and a dense cloud in the external medium \citep[e.g., as in 3C\,120;][]{Gomez_2000,Gomez_2001}, although the lack of clear signatures of jet disruption render this interpretation less likely. 
Instabilities in a relativistic jet constitute another possible origin of the bent structure. There are two main types of instabilities that can be responsible for this bending: (1) the Kelvin-Helmholtz (KH) instability; and (2) the current-driven (CD) kink instability. KH instabilities develop through the shear between two different flow components, e.g., the fast jet spine and the slow jet sheath and/or the wind and external medium \citep{Mizuno2007, Sironi_2021}. The non-axisymmetric helical mode can produce bent structures in relativistic jets \citep{Hardee2000, Lobanov2001}. This instability grows in the kinetic energy dominated region; therefore the region far from the jet base is a preferred site. The CD kink instability was also shown to generate helically twisted jet structures \citep{McKinney2009,Mizuno2012,Davelaar_2020}. This instability is excited by the existence of a helical magnetic field, which is predicted by the jet formation theory and simulations \citep[e.g.,][]{Pudritz2012}, and is expected to grow in the magnetically dominated region near the jet base. The toroidal magnetic fields would create a twisted polarization pattern like the one we observe in the $P0$ component at the VLBI core in Figure~\ref{fig:pol_images}.

\subsection{Brightness temperatures}

We measure the observer frame brightness temperatures $T_{\rm B}$ of the VLBI core by performing Gaussian model fitting in  {\tt DIFMAP} at each frequency, and estimating the core size and flux density. The results are presented in Table~\ref{tab:Tb}, with $T_{\rm B}$ calculated as
\begin{equation}
    T_{\rm B} = 1.22 \times 10^{12} \frac{F_\nu}{\nu^2 \theta^2} \ ,
\end{equation} 
where the units of $F_\nu$, $\nu$, and $\theta$ are as given in Table \ref{tab:Tb}. Apart from that, non-simultaneous MOJAVE data at 15\,GHz give a core brightness temperature $\sim 10^{12}$\,K in 2012, and an apparent brightness temperature in excess of $10^{12}$\,K was found at 1.6\,GHz \citep{Shen_1999}.
\begin{table}[htp]
\caption{Properties of the VLBI core at each observing frequency from {\tt DIFMAP} Gaussian component fitting.}
\begin{center}
\tabcolsep=0.25cm
\begin{tabularx}{0.8\linewidth}{ccccc}
\hline
\hline
$\nu$ (GHz) & 2.3 & 8.7 & 86.2 & 229.1 \\
$F_\nu$ (Jy) & 2.2 & 4.3 & 0.4 & 0.5 \\  
  $\theta$ (mas) & 0.82 &  0.56 &  --$^{a)}$ &  0.01 \\
  $T_{\rm B}$ $\left(10^{11}\,{\rm K} \right)$ & 7.6 &  7.8 &  2.8 &  1.2 \\
 \hline 
\end{tabularx}
\end{center}
$^{a)}$ a $15\,\mu$as resolution limit calculated using the residual map noise was used
\label{tab:Tb}
\end{table}
The reported brightness temperatures are generally lower limits if the cores are not resolved. Nevertheless, the trend of brightness temperature decreasing with frequency seems robust, and consistent with signatures of an accelerating (sub-)pc scale jet \citep{Lee2016}. However, this interpretation is only straightforward if the change in inclination angle in the bent jet does not cause significant changes in the Doppler factor. The 230\,GHz brightness temperature is consistent with the visibility-domain brightness temperature limit obtained from the flux density values observed on the longest baselines \citep{Lobanov2015}, which, in our case, reach about 8.5\,G$\lambda$, as well as with the prior estimates by \citet{Lu2012}. The observed value is related to the fluid frame brightness temperature $T^{'}_{\rm B}$ via
\begin{equation}
    T^{'}_{\rm B} = T_{\rm B} \frac{1+z}{\delta} 
\end{equation}
with a Doppler factor $\delta$ and Lorentz factor $\Gamma$
\begin{equation}
    \delta = \frac{1}{\Gamma (1 - \beta \cos \theta) } \ ; \ \Gamma = (1-\beta^2)^{-1/2} \ 
\end{equation}
with velocity $\beta$ and viewing angle $\theta$. While we do not have strong limits on the jet velocity and inclination 
(which also varies with the angular scale as the jet is bent), \citet{Paliya_2017} suggests $\Gamma=12$ ($\beta > 0.996\,c$), and it is enough that $\theta < 20$\,deg and $\beta > 0.6\,c$ in order to obtain $\delta > 2$. This implies that the intrinsic brightness temperature at 230\,GHz is likely lower than the equipartition temperature $T_{\rm eq}$, $T^{'}_{\rm B} < T_{\rm eq} = 5 \times 10^{10}$\,K, indicating a magnetically-dominated inner jet \citep{Redhead1994}, likely dominated by the helical magnetic field. The brightness temperature $T^{'}_{\rm B}$ is also significantly lower than the inverse-Compton limit of $\sim 5 \times 10^{11}$\,K \citep{Kellermann1969}. 
These findings are comparable to the ones reported for 3C\,279 by \citet{Kim2020} from observations in the same EHT run.
Low brightness temperatures at 230\,GHz were also reported for other EHT sources, Cen~A \citep[resolved at about 200 $R_{\rm S}$ scale,][]{Janssen2021} and M\,87$^*$ \citep[resolved at about 3.5 $R_{\rm S}$ scale,][]{PaperI}.

\subsection{Superresolved millimeter core} 
\label{sec:core}

With RML-based imaging methods we may expect to achieve superresolution in regularized fitted images \citep{Honma2014}. Indeed, with \texttt{eht-imaging} we consistently find the $C0$ component as an elliptical feature of major and minor axes full widths at half maximum equal to 15\,$\mu$as and 10\,$\mu$as, respectively, and a major axis PA of about 45 deg, perpendicular to the position angle of $C1$, which we associate with the PA of the sub-pc scale jet. Superresolved images obtained with an RML-based reconstruction method, with no subsequent blurring, are shown in Appendix~\ref{app:hops_casa}. In all images in Figure \ref{fig:pol_images}, the polarized component $P0$ appears to have a crescent-like shape around the core $C0$, with a large fractional polarization of about 15\% and a depolarized intrusion on the west side of the core feature. Within the $P0$ structure, the EVPA rotates by at least 90 deg around the core, consistently across days and methods, see Figure~\ref{fig:pol_images}. We interpret this as a signature of the presence of toroidal magnetic fields in the core \citep{Molina2014}. The inner depolarization could be a resolution effect related to the size of the beam, averaging over the spatially varying EVPA within the VLBI core region, however the depolarization towards the west requires a different explanation. It is possible that the innermost jet is launched in the western direction, and hence because the material in the jet is partially obscuring the view onto the jet base, the Faraday depth is greater on the west side of the core, causing depolarization. The western orientation of the inner jet 
is consistent with the trend of the bend direction seen across all angular scales. An alternative explanation is a conical shock \citep{Lind1985}, which can reproduce similar geometrical features in the EVPA orientation, particularly in the presence of a tangled magnetic field component \citep{Cawthorne2006}. In this interpretation the core in our images would need to be associated with the optically thin jet plasma. 

Simultaneous interferometric-ALMA observations of \janine give spectral indices of $-0.5 \pm 0.10$ at 93\,GHz, and $-0.75 \pm 0.10$ at 220\,GHz \citep{Goddi2021}. These values suggest an optically thin source, and since the emission is dominated by the core component, this could imply an optically thin core, contrary to the standard model of a radio jet, in which the core emission corresponds to the photosphere of the optically thick region \citep{Blandford_Konigl_1979}. Such optically thin emission is expected if the source is observed at a frequency higher than the synchrotron self-absorption turnover. In this case the VLBI core would appear resolved, and the polarization substructure would become observable.

\subsection{Jet features}
\label{sec:jet_features}

The two total intensity jet features $C1$ and $C2$ appear at PAs of -47 deg and -36 deg respectively, with a change consistent with the direction of the jet bending. Additionally, we find a polarized component $P1$, located between $C1$ and $C2$. The EVPA in the $P1$ component is ordered and aligned in a pattern parallel to the jet. This is indicative of a transverse magnetic field component, implying a toroidal or helical magnetic field topology, and this provides a simple and natural explanation for the presence of the $P1$ offset from $C1$ and $C2$. It is possible that $C2$ is a relativistic transverse shock in the jet, which enhances the magnetic field in the plane of compression, perpendicularly to the shock propagation direction \citep{Hughes1985}. In this case, some of the polarization associated with $P1$ could be associated with this shock structure. The separation between the linearly polarized and total intensity features could be a consequence of the presence of sub-structure within the shock, e.g., a forward and reverse shock, one being more polarized than the other \citep{Gomez1997,Laskar2019}.
There is some weak indication of $P1$ and $C2$ motion in opposite directions supporting this interpretation, see Figure~\ref{fig:featuresB}. Linear polarization maps in 15\,GHz with MOJAVE indicate that in the most upstream region the EVPA is aligned with the 230\,GHz jet PA \citep{Lister2018}. This suggests a similar origin of the 15\,GHz polarization features as transverse shock features in the upstream jet, unresolved at 15\,GHz. Further away from the core, the 15\,GHz maps show the EVPA aligning with the mas scale jet orientation, strengthening this interpretation.

\section{Summary}\label{sec:summary}

In this paper, we presented the first 1.3\,mm VLBI total-intensity and polarimetric images of the blazar \janine with the EHT. The EHT enabled the highest resolution polarimetric imaging of a quasar to date, corresponding to a linear resolution of $\sim$0.1\,pc. These unprecedented images of the inner parsec of \janine reveal a compact total-intensity structure of three distinct components oriented in a north-west direction, with a fan-like EVPA pattern in the VLBI core (the southernmost component). We did not find significant motion of the component $C1$, closest to the core, with an upper limit of 2\,$c$ on the apparent velocity. In the superresolved core region, 
we notice a rotation of the EVPA, suggestive of the presence of toroidal magnetic fields in the core region.
We have shown that \janine is a bright and very compact source at mm wavelengths, displaying very little variability on a timescale of several days -- these features render it possibly the best available EHT calibrator positioned close to Sgr\,A$^*$ on the sky.

We compared our EHT images with quasi-simultaneous images of \janine at longer wavelengths obtained with the GMVA and the VLBA. We observed a clockwise rotation of the jet direction in \janine as we go from long to short observing wavelengths, with an apparent bend of the jet in 3.5\,mm. The rotation of the PA with the frequency could be indicative of a helical jet structure. Several scenarios have been proposed for helical jets in other sources, such as a putative supermassive black hole binary in the core, or a tilted large-scale accretion flow compared to the black hole spin axis, or shock regions as the jet interacts with the external medium. All these scenarios predict a time variability of the PA at individual frequencies, which we do not see in the $\sim$20 year timescale of the MOJAVE 15\,GHz observations on angular scales of $\sim\,5$\,mas. Monitoring on longer timescales, particularly at higher frequencies, will be needed to further understand the helical structure. 

The narrow fractional bandwidth of the EHT 2017 observations ($\Delta \nu / \nu < 2\%$), and a scale separation between observations by different arrays prevented us from studying the spatially resolved spectral index and rotation measure. EHT observations in 2018 and later provide a wider bandwidth ($\Delta \nu / \nu > 6\%$), alleviating this shortcoming \citep{PaperII}. There are plans for expanding the EHT array and enabling 345\,GHz observations \citep{Doeleman2019}, which should further improve both the resolution and the dynamic range of the \janine images.

Finally, \janine is a source of $\gamma$-ray radiation identified in the Fermi-LAT catalog \citep{Fermi2020}. Given that with the EHT at 1.3\,mm we see detailed structure of the source total intensity and linear polarization on an extreme scale of $\sim 0.1$\,pc, the source may be an excellent target to study the relation between high energy emission and jet morphology and kinematics at millimeter wavelengths. Interestingly, the most recent 230\,GHz SMA monitoring data from the beginning of 2022 show a steep brightness rise to the largest values seen in a decade. 

\section{Acknowledgments}

We thank the anonymous reviewer for their thoughtful and helpful comments.
The Event Horizon Telescope Collaboration thanks the following
organizations and programs: the Academy
of Finland (projects 274477, 284495, 312496, 315721); the Agencia Nacional de Investigación y Desarrollo (ANID), Chile via NCN$19\_058$ (TITANs) and Fondecyt 3190878, the Alexander
von Humboldt Stiftung; an Alfred P. Sloan Research Fellowship;
Allegro, the European ALMA Regional Centre node in the Netherlands, the NL astronomy
research network NOVA and the astronomy institutes of the University of Amsterdam, Leiden University and Radboud University;
the Black Hole Initiative at
Harvard University, through a grant (60477) from
the John Templeton Foundation; the China Scholarship
Council;  Consejo
Nacional de Ciencia y Tecnolog\'{\i}a (CONACYT,
Mexico, projects  U0004-246083, U0004-259839, F0003-272050, M0037-279006, F0003-281692,
104497, 275201, 263356);
the Delaney Family via the Delaney Family John A.
Wheeler Chair at Perimeter Institute; Dirección General
de Asuntos del Personal Académico-—Universidad
Nacional Autónoma de México (DGAPA-—UNAM,
projects IN112417 and IN112820); the European Research Council Synergy
Grant "BlackHoleCam: Imaging the Event Horizon
of Black Holes" (grant 610058); the Generalitat
Valenciana postdoctoral grant APOSTD/2018/177 and
GenT Program (project CIDEGENT/2018/021); MICINN Research Project PID2019-108995GB-C22;
the
Gordon and Betty Moore Foundation (grant GBMF-3561); the Istituto Nazionale di Fisica
Nucleare (INFN) sezione di Napoli, iniziative specifiche
TEONGRAV; the International Max Planck Research
School for Astronomy and Astrophysics at the
Universities of Bonn and Cologne; 
Joint Princeton/Flatiron and Joint Columbia/Flatiron Postdoctoral Fellowships, research at the Flatiron Institute is supported by the Simons Foundation; 
the Japanese Government (Monbukagakusho:
MEXT) Scholarship; the Japan Society for
the Promotion of Science (JSPS) Grant-in-Aid for JSPS
Research Fellowship (JP17J08829); the Key Research
Program of Frontier Sciences, Chinese Academy of
Sciences (CAS, grants QYZDJ-SSW-SLH057, QYZDJSSW-
SYS008, ZDBS-LY-SLH011); the Leverhulme Trust Early Career Research
Fellowship; the Max-Planck-Gesellschaft (MPG);
the Max Planck Partner Group of the MPG and the
CAS; the MEXT/JSPS KAKENHI (grants 18KK0090,
JP18K13594, JP18K03656, JP18H03721, 18K03709,
18H01245, 25120007); the Malaysian Fundamental Research Grant Scheme (FRGS) FRGS/1/2019/STG02/UM/02/6; the MIT International Science
and Technology Initiatives (MISTI) Funds; the Ministry
of Science and Technology (MOST) of Taiwan (105-
2112-M-001-025-MY3, 106-2112-M-001-011, 106-2119-
M-001-027, 107-2119-M-001-017, 107-2119-M-001-020,
107-2119-M-110-005, 108-2112-M-001-048, and 109-2124-M-001-005); the National Aeronautics and
Space Administration (NASA, Fermi Guest Investigator
grant 80NSSC20K1567, NASA Astrophysics Theory Program grant 80NSSC20K0527, NASA NuSTAR award 80NSSC20K0645); the National
Institute of Natural Sciences (NINS) of Japan; the National
Key Research and Development Program of China
(grant 2016YFA0400704, 2016YFA0400702); the National
Science Foundation (NSF, grants AST-0096454,
AST-0352953, AST-0521233, AST-0705062, AST-0905844, AST-0922984, AST-1126433, AST-1140030,
DGE-1144085, AST-1207704, AST-1207730, AST-1207752, MRI-1228509, OPP-1248097, AST-1310896,  
AST-1555365, AST-1615796, AST-1715061, AST-1716327,  AST-1903847,AST-2034306); the Natural Science
Foundation of China (grants 11573051, 11633006,
11650110427, 10625314, 11721303, 11725312, 11933007, 11991052, 11991053); a
fellowship of China Postdoctoral Science Foundation (2020M671266); the Natural
Sciences and Engineering Research Council of
Canada (NSERC, including a Discovery Grant and
the NSERC Alexander Graham Bell Canada Graduate
Scholarships-Doctoral Program); the National Youth
Thousand Talents Program of China; the National Research
Foundation of Korea (the Global PhD Fellowship
Grant: grants NRF-2015H1A2A1033752, 2015-
R1D1A1A01056807, the Korea Research Fellowship Program:
NRF-2015H1D3A1066561, Basic Research Support Grant 2019R1F1A1059721); the Netherlands Organization
for Scientific Research (NWO) VICI award
(grant 639.043.513) and Spinoza Prize SPI 78-409; the
New Scientific Frontiers with Precision Radio Interferometry
Fellowship awarded by the South African Radio
Astronomy Observatory (SARAO), which is a facility
of the National Research Foundation (NRF), an
agency of the Department of Science and Technology
(DST) of South Africa; the Onsala Space Observatory
(OSO) national infrastructure, for the provisioning
of its facilities/observational support (OSO receives
funding through the Swedish Research Council under
grant 2017-00648) the Perimeter Institute for Theoretical
Physics (research at Perimeter Institute is supported
by the Government of Canada through the Department
of Innovation, Science and Economic Development
and by the Province of Ontario through the
Ministry of Research, Innovation and Science);  the Spanish
Ministerio de Economía y Competitividad (grants
PGC2018-098915-B-C21, AYA2016-80889-P, PID2019-108995GB-C21); the State
Agency for Research of the Spanish MCIU through
the "Center of Excellence Severo Ochoa" award for
the Instituto de Astrofísica de Andalucía (SEV-2017-
0709); the Toray Science Foundation; the Consejería de Economía, Conocimiento, Empresas y Universidad 
of the Junta de Andalucía (grant P18-FR-1769), the Consejo Superior de Investigaciones Científicas (grant 2019AEP112);
the US Department
of Energy (USDOE) through the Los Alamos National
Laboratory (operated by Triad National Security,
LLC, for the National Nuclear Security Administration
of the USDOE (Contract 89233218CNA000001);
 the European Union’s Horizon 2020
research and innovation programme under grant agreement
No 730562 RadioNet; ALMA North America Development
Fund; the Academia Sinica; Chandra DD7-18089X and TM6-
17006X; the GenT Program (Generalitat Valenciana)
Project CIDEGENT/2018/021. This work used the
Extreme Science and Engineering Discovery Environment
(XSEDE), supported by NSF grant ACI-1548562,
and CyVerse, supported by NSF grants DBI-0735191,
DBI-1265383, and DBI-1743442. XSEDE Stampede2 resource
at TACC was allocated through TG-AST170024
and TG-AST080026N. XSEDE JetStream resource at
PTI and TACC was allocated through AST170028.
The simulations were performed in part on the SuperMUC
cluster at the LRZ in Garching, on the
LOEWE cluster in CSC in Frankfurt, and on the
HazelHen cluster at the HLRS in Stuttgart. This
research was enabled in part by support provided
by Compute Ontario (http://computeontario.ca), Calcul
Quebec (http://www.calculquebec.ca) and Compute
Canada (http://www.computecanada.ca). We thank
the staff at the participating observatories, correlation
centers, and institutions for their enthusiastic support. This paper makes use of the following ALMA data:
ADS/JAO.ALMA\#2016.1.01154.V and ADS/JAO.ALMA2016.1.00413.V. ALMA is a partnership of the European Southern Observatory (ESO;
Europe, representing its member states), NSF, and
National Institutes of Natural Sciences of Japan, together
with National Research Council (Canada), Ministry
of Science and Technology (MOST; Taiwan),
Academia Sinica Institute of Astronomy and Astrophysics
(ASIAA; Taiwan), and Korea Astronomy and
Space Science Institute (KASI; Republic of Korea), in
cooperation with the Republic of Chile. The Joint
ALMA Observatory is operated by ESO, Associated
Universities, Inc. (AUI)/NRAO, and the National Astronomical
Observatory of Japan (NAOJ). The NRAO
is a facility of the NSF operated under cooperative agreement
by AUI. APEX is a collaboration between the
Max-Planck-Institut f{\"u}r Radioastronomie (Germany),
ESO, and the Onsala Space Observatory (Sweden). The
SMA is a joint project between the SAO and ASIAA
and is funded by the Smithsonian Institution and the
Academia Sinica. The JCMT is operated by the East
Asian Observatory on behalf of the NAOJ, ASIAA, and
KASI, as well as the Ministry of Finance of China, Chinese
Academy of Sciences, and the National Key R\&D
Program (No. 2017YFA0402700) of China. Additional
funding support for the JCMT is provided by the Science
and Technologies Facility Council (UK) and participating
universities in the UK and Canada. The
LMT is a project operated by the Instituto Nacional
de Astrófisica, Óptica, y Electrónica (Mexico) and the
University of Massachusetts at Amherst (USA). The
IRAM 30-m telescope on Pico Veleta, Spain is operated
by IRAM and supported by CNRS (Centre National de
la Recherche Scientifique, France), MPG (Max-Planck-
Gesellschaft, Germany) and IGN (Instituto Geográfico
Nacional, Spain). The SMT is operated by the Arizona
Radio Observatory, a part of the Steward Observatory
of the University of Arizona, with financial support of
operations from the State of Arizona and financial support
for instrumentation development from the NSF.
Support for SPT participation in the EHT is provided by 
the National Science Foundation through award OPP-1852617 
to the University of Chicago. Partial support is also 
provided by the Kavli Institute of Cosmological Physics 
at the University of Chicago. The SPT hydrogen maser was 
provided on loan from the GLT, courtesy of ASIAA.
The EHTC has received generous donations of FPGA chips from Xilinx
Inc., under the Xilinx University Program. The EHTC
has benefited from technology shared under open-source
license by the Collaboration for Astronomy Signal Processing
and Electronics Research (CASPER). The EHT
project is grateful to T4Science and Microsemi for their
assistance with Hydrogen Masers. This research has
made use of NASA’s Astrophysics Data System. We
gratefully acknowledge the support provided by the extended
staff of the ALMA, both from the inception of
the ALMA Phasing Project through the observational
campaigns of 2017 and 2018. We would like to thank
A. Deller and W. Brisken for EHT-specific support with
the use of DiFX. We acknowledge the significance that
Mauna Kea, where the SMA and JCMT EHT stations
are located, has for the indigenous Hawaiian people. 

We also thank Alexandra Elbakyan for her contributions to the open science initiative.  This research has made use of data obtained with the Global Millimeter VLBI Array (GMVA), coordinated by the VLBI group at the Max-Planck-Institut f{\"u}r Radioastronomie (MPIfR). The GMVA consists of telescopes operated by MPIfR, IRAM, Onsala, Metsahovi, Yebes, the Korean VLBI Network, the Green Bank Observatory and the Very Long Baseline Array (VLBA). The VLBA and the GBT are facilities of the National Science Foundation under cooperative agreement by Associated Universities, Inc. The data were correlated at the DiFX correlator of the MPIfR in Bonn, Germany. We thank the National Science Foundation (awards OISE-1743747, AST-1816420, AST-1716536, AST-1440254, AST-1935980) and the Gordon and Betty Moore Foundation (GBMF-5278) for financial support of this work. Support for this work was also provided by the NASA Hubble Fellowship grant
HST-HF2-51431.001-A awarded by the Space Telescope
Science Institute, which is operated by the Association
of Universities for Research in Astronomy, Inc.,
for NASA, under contract NAS5-26555.

\appendix

\section{Data reduction pipelines comparison}
\label{app:hops_casa}
\begin{figure}[h]
    \centering
    \includegraphics[width=0.999\columnwidth]{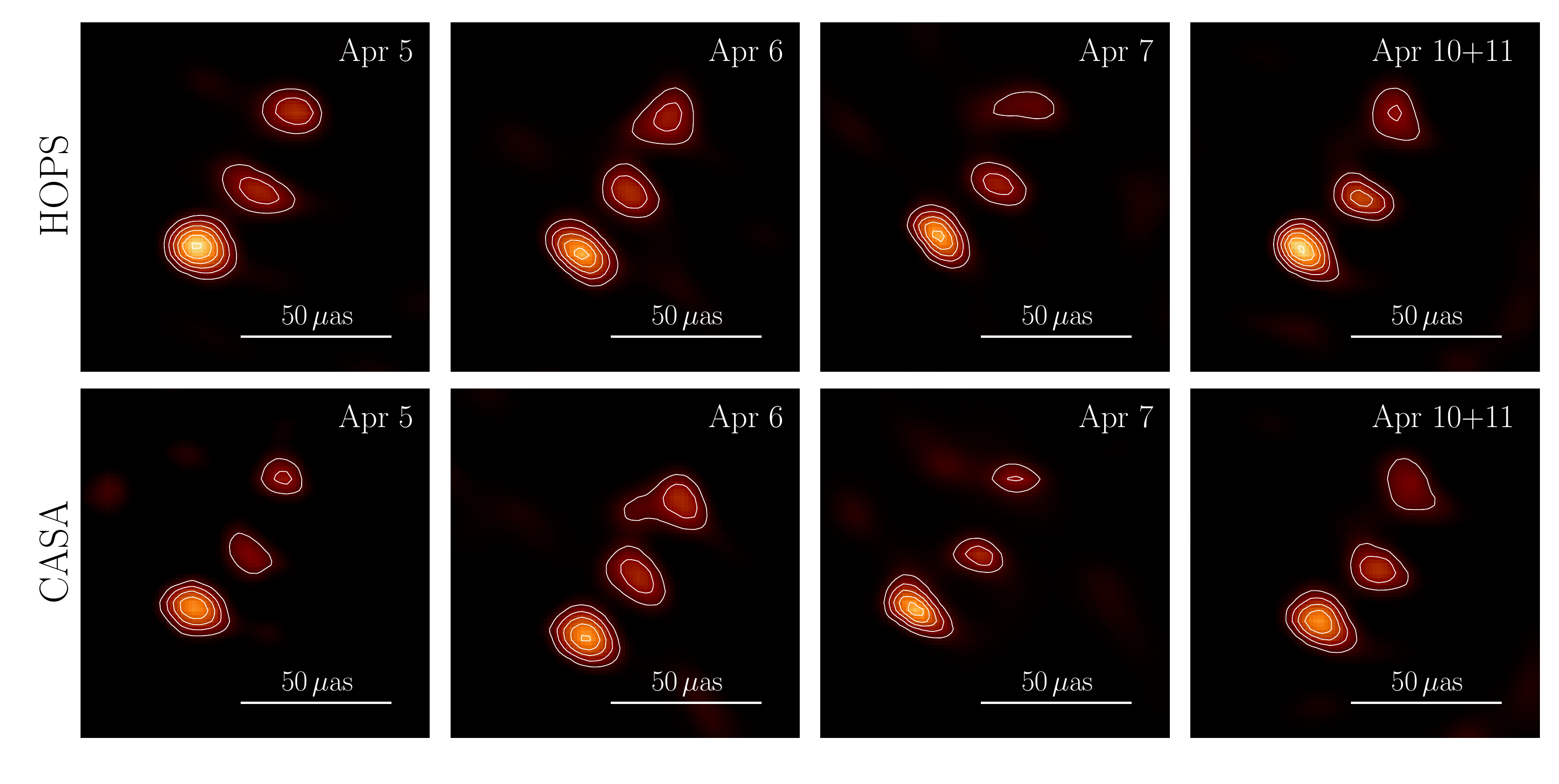}
    \caption{Total intensity images of \janine for the four EHT observing epochs in April 2017, for two independent data calibration pipelines {\tt EHT-HOPS} (upper panels) and {\tt CASA-rPICARD} (bottom panels). The color scale is the same for all images, with a peak value corresponding to 10\,mJy/$\mu$as$^2$. Contours correspond to 15, 30, 45, 60, 75, 90\% of the peak flux density. There is a high level of consistency between the two calibration methods.}
    \label{fig:hops_casa}
\end{figure}

In Figure \ref{fig:hops_casa} we compare the Stokes $\mathcal{I}$ (total intensity) 230\,GHz images obtained from the two data reduction pipelines: {\tt EHT-HOPS} \citep{Blackburn_2019}, and {\tt CASA}-based {\tt rPICARD} \citep{Janssen2019}. All images presented in Figure \ref{fig:hops_casa} were obtained using an identical {\tt eht-imaging} script \citep{Chael_2016}. The images correspond to a direct fit to the observations, with no blurring or restoring beam applied.

\section{Leakage terms consistency}
\label{app:leakage}

\begin{figure}[h]
    \centering
    \includegraphics[width=0.999\columnwidth]{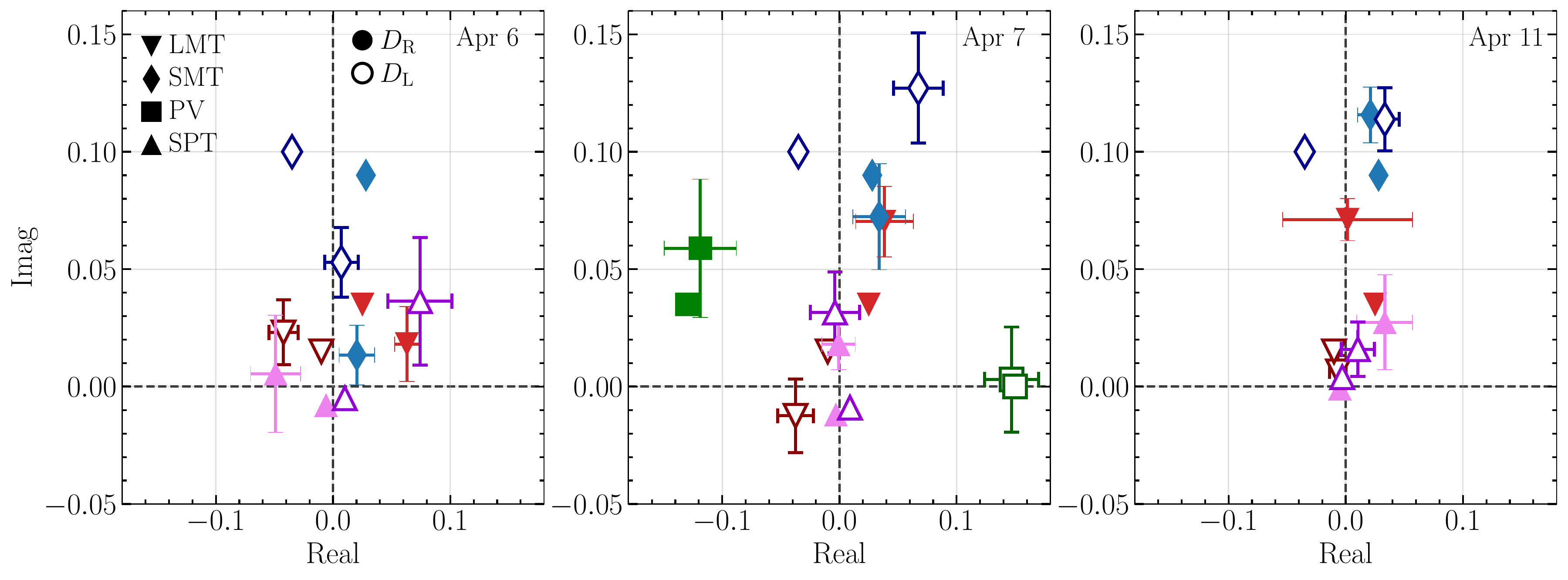}
    \caption{Leakage terms estimated with \texttt{DMC} for \janine data for 2017 April 6, 7, and 11, shown with 1\,$\sigma$ errorbars. The matching symbols without errorbars correspond to the assumed leakage terms given in Table \ref{tab:dterms}. Filled and open symbols denote right and left circular polarization D-terms, respectively.
 In case of the SPT, markers correspond to leakage terms estimated by \texttt{eht-imaging}, separately for each observing day. 
 }
    \label{fig:Dterms}
\end{figure}

Leakage coefficients can be estimated robustly for the telescopes with an intra-site station (ALMA, APEX, SMA, JCMT), where a point source model can be employed for a multi-source fit \citep{Marti2021,PaperVII}. For the remaining sites the leakage terms need to be modeled simultaneously with the source structure and the problem becomes degenerate. The values presented in Table~\ref{tab:dterms} are representative values based on the analysis spanning multiple epochs and different image reconstruction algorithms \citep{PaperVII}. These values were used for the linear polarization imaging with \texttt{eht-imaging}, and only SPT D-terms were solved for in this framework. For imaging with \texttt{DMC}, leakage terms were fitted within the software, providing a consistency test with the assumed values, see Figure~\ref{fig:Dterms}. Systematic differences between days are seen for the \texttt{DMC} fits. Since D-terms are unlikely to significantly vary in time for stations other than ALMA \citep{Goddi2019}, this highlights the importance of using multi-epoch fits to constrain the leakage coefficients. SPT leakage calibration is particularly challenging given the limited parallactic coverage and the relevant D-terms have not been estimated in \citet{PaperVII} because M\,87$^*$ is not observable from the south pole. Our fits to \janine data indicate that the magnitude of the SPT D-terms does not exceed 5\%. 
Overall the D-terms have an acceptable degree of consistency and the residual uncertainties related to imperfect leakage calibration do not influence the overall morphology of the linear polarization images, see Section~\ref{sec:linear_polarization}.

\bibliographystyle{aa}
\bibliography{J1924.bib}

\end{document}